\documentclass{iopjournal}

\usepackage[english]{babel}

\usepackage[utf8]{inputenc} 
\usepackage[T1]{fontenc}
\usepackage{textcomp}

\usepackage{amsthm}
\usepackage{amsfonts}
\usepackage{amssymb}
\usepackage{amsbsy}

\usepackage[squaren,Gray]{SIunits}

\usepackage{graphicx}
\usepackage{placeins}
\usepackage{tabularx}
\usepackage{xcolor}
\usepackage{float}

\usepackage{caption}
\usepackage{subcaption}
\usepackage{empheq}

\usepackage{comment}
\usepackage{keyval}
\usepackage{hyperref}
\usepackage{cite}
\usepackage{orcidlink}

\usepackage{afterpage}

\numberwithin{equation}{section}

\newcommand\Grad{\Vec{\nabla}}

\newcommand\B{\Vec{B}}

\providecommand{\keywords}[1]
{
  \small	
  \textbf{\textit{Keywords---}} #1
}

\begin{document}

\begingroup
\let\clearpage\relax
\twocolumn[
\begin{@twocolumnfalse}
\title{Kinetic Equilibrium Prediction at TCV using RAPTOR and FBT}

\author{
C.~E.~Contré$^{1}$\orcid{0009-0008-3246-1860},
A.~Merle$^{1}$\orcid{0000-0003-1831-5644},
O.~Sauter$^{1}$\orcid{0000-0002-0099-6675},
S.~Van~Mulders$^{1,2}$\orcid{0000-0003-3184-3361},
R.~Coosemans$^{1}$\orcid{0000-0001-8110-3156},
G.~Durr-Legoupil-Nicoud$^{1}$\orcid{0009-0002-5956-6482},
F.~Felici$^{1}$\orcid{0000-0001-7585-376X},
O.~Février$^{1}$\orcid{0000-0002-9290-7413},
C.~Heiss$^{1}$\orcid{0009-0005-9712-0643},
B.~Labit$^{1}$\orcid{0000-0002-0751-8182},
A.~Pau$^{1}$\orcid{0000-0002-7122-3346},
Y.~Poels$^{1,3}$\orcid{0000-0002-4071-4855},
C.~Venturini$^{1}$\orcid{0009-0005-9873-1171},
B.~Vincent$^{1}$\orcid{0000-0001-5420-6002},
the TCV team$^{4}$,
and the EUROfusion Tokamak Exploitation Team$^{5}$
}

\affil{$^{1}$Ecole Polytechnique Fédérale de Lausanne (EPFL), Swiss Plasma Center (SPC), CH-1015 Lausanne, Switzerland\\
$^{2}$ITER Organization, Route de Vinon sur Verdon, St Paul Lez Durance, 13115, France\\
$^{3}$Eindhoven University of Technology, Mathematics and Computer Science, NL-5600MB Eindhoven, The Netherlands\\
$^{4}$See author list of B. P. Duval et al. 2024 Nucl. Fusion 64 112023\\
$^{5}$See author list of E. Joffrin et al. 2024 Nucl. Fusion 64 112019
}

\email{cassandre.contre@epfl.ch}

\keywords{RAPTOR, FBT, KEP, Equilibrium, Transport, Predict-first, TCV}

\begin{abstract}
We present results from a new Kinetic-Equilibrium Prediction (KEP) workflow and shot preparation for full TCV discharges, by coupling predict-first RAPTOR transport simulations with FBT inverse equilibrium calculations. RAPTOR is a 1.5D transport code which has been extensively used for plasma shot optimization and real-time modeling. We show that rapid pre-shot simulations can be performed directly using information from the pulse schedule across a wide range of plasma shapes and scenarios, given an estimate of the confinement quality factor $H^{98(y,2)}$ and line-averaged density. The resulting $p'$ and $TT'$ profiles are then provided to the pre-shot equilibrium computation performed by FBT --- a static free-boundary solver routinely used at TCV --- achieving convergence between the two codes in a few iterations. Finally, we show that this coupling, when integrated into the TCV shot preparation, improves the evaluation of the coil currents needed to match the target plasma shape; in particular providing an accurate estimate of critical quantities such as the internal inductance $l_i$ and normalized pressure $\beta_N$, giving more realistic information to tokamak operators about the expected pulse behavior and enabling them to adjust the plan correspondingly. 

\end{abstract}

\end{@twocolumnfalse}
]
\endgroup

\section{Introduction}
Any attempt to optimize tokamak performance is inevitably confronted to operational limits, beyond which machine protection systems or disruptions may prematurely terminate a pulse \cite{ciattaglia_iter_2009}. Most of the 0D limits known to date to prevent these events have been formalized since decades \cite{hender_chapter_2007,bandyopadhyay_mhd_2025}, although regularly refined \cite{giacomin_first-principles_2022}, and are now increasingly compared with results from integrated pulse simulation workflows, to assess scenario feasibility in light of the full non-linear spatial and temporal dynamics of a plasma discharge. In particular, pulse design and optimization tools, able to simulate, optimize and prepare a scenario shortly before an experiment \cite{artaud_metis_2018,lyons_flexible_2023,luda_validation_2021,romanelli_jintrac_2014}, and flight simulators, which integrate the response of the actuators to the control system \cite{janky_validation_2021}, are two complementary tools used in shot preparation which demand to simulate a whole plasma discharge, from ramp-up to ramp-down, with reasonable computational time. Modeling scenarios with these tools will be a prerequisite for conducting experiments on next step devices like ITER \cite{hender_chapter_2007}, while smaller tokamaks like the \textit{Tokamak à Configuration Variable} (TCV) \cite{Hofmann_1994} can be used for testing and validation. 
In this paper, we present the development of a predict-first pulse planning workflow in which RAPTOR \cite{felici_real-time_2011} is used to predict the self-consistent response of the current density, electron and ion temperatures, and particle density profiles to actuators, while providing these profiles to FBT \cite{hofmann_fbt_1988} for computing the equilibrium and coil currents required to sustain the desired shape and position. RAPTOR is a rapid solver of the 1D radial transport equations in the plasma core. It is implemented in MATLAB with the ability to automatically generate code in C, for accelerating computations or integration in real-time control systems, for example in
 TCV \cite{felici_real-time_2011,felici_real-time-capable_2018,carpanese_first_2020}, RFX \cite{piron_integration_2017}, JET \cite{piron_development_2021} and ASDEX-Upgrade \cite{kudlacek_overview_2024, REISNER2026115474}. Previous works also demonstrated its ability to optimize scenarios \cite{teplukhina_simulation_2017,van_mulders_rapid_2021,van_mulders_scenario_2023,van_mulders_scenario_2023-1,van_mulders_inter-discharge_2024} and to integrate new physics models, such as the Qualikiz surrogate model, QLKNN-10D \cite{van_de_plassche_fast_2020,van_mulders_rapid_2021}. However, its ability to predict transport rapidly and automatically before an experiment remained to be demonstrated. 
By using the gradient-based transport model \cite{teplukhina_simulation_2017,kim_simple_2016} together with the Martin scaling law \cite{martin_power_2008} for predicting the L- to H-mode transition times, this work investigates the extent to which it is possible to predict with RAPTOR the evolution of the kinetic and current density profiles, including the evolution of the poloidal beta $\beta_{pol}$ and internal inductance $l_i$, from the current ramp-up to ramp-down, purely based on information available before the shot. We will also show, in particular, that estimating the line-averaged density ($n_{e,l}$) from the gas programming remains a delicate and essential step towards predicting the kinetic profiles before the discharge, and would need further investigation. Similarly, improving our ability to predict L-H/H-L transitions for various scenarios is also an important step. Therefore, a validation over a large TCV database is presented, similarly to what has been done in \cite{abbate_large-database_2024} for comparing post-shot ASTRA and TRANSP simulations with DIII-D data. In \cite{wan_machine_2026} and references therein, machine learning technics have been applied to rapidly predict tokamak discharges, using both poloidal field coil currents and line-averaged density as inputs from experiments. Similar efforts of applying AI-based models to the pulse prediction have also been reported for the TJ-II stellerator in \cite{bustos_ai-based_2025}, taking magnetic fluctuations as inputs from experiments and predicting the density fuelling and heating configuration from the actuator programming. Rapidly predicting the current diffusion and heat and particle transport in the plasma core aims to inform scientific coordinators on the evolution of key physical quantities for scenario planning, but also to inform other modeling tools. 
\begin{figure}[t]
    \centering
    \includegraphics[width=\linewidth]{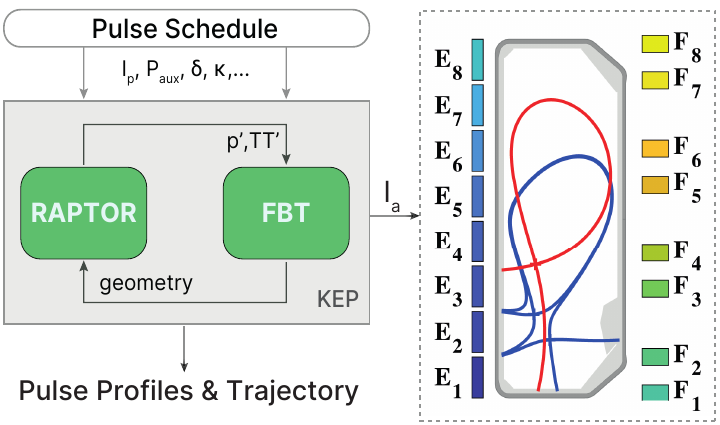}
    \vspace{0.05em}
    \captionsetup{width=\linewidth}
    \caption{Schematic overview of the Kinetic Equilibrium Prediction (KEP) workflow. The pulse schedule is given as input to RAPTOR and FBT, loosely coupled to provide the active coil currents and pulse profiles predictively. These new feedforward currents traces are then integrated into the TCV preparation system to control the plasma shape and position with its 16 separately powered PF coils. Examples of $\#82030$ (lower single-null, LSN, in red) and $\#83575$ (snowflake (SF) upper NT, in blue) shapes, achievable at TCV and simulated with the KEP (see Sec. \ref{Sec:TCV_results}), are also represented.}
    \label{fig:legend_coils_polview}
\end{figure}

Predicting plasma pressure and current density with RAPTOR can indeed help to improve the correctness of the equilibrium used to design the poloidal field (PF) coil currents by providing an estimate of the pressure gradient $p'(\psi, t)$ and current function $TT'(\psi, t)$ ($T=RB_\phi$) time evolution. How to provide realistic internal profiles to equilibrium calculations is a longstanding challenge. In reconstruction codes, like LIUQE \cite{moret_tokamak_2015}, this problem can be solved by constraining the profiles with experimental measurements and modeling. This coupling method, called Kinetic Equilibrium Reconstruction (KER), has already been demonstrated with RAPTOR and LIUQE for real-time and post-shot reconstruction in \cite{carpanese_first_2020} and \cite{van_mulders_model-based_2026}, as also done in the ASTRA-LIUQE KER \cite{carpanese_development_2021}, CLISTE  \cite{mc_carthy_analytical_1999}, IDE \cite{fischer_integrated_2010,fischer_estimation_2020} and EFIT  \cite{lao_reconstruction_1985,xing_cake_2021}. In a similar way, preparing the equilibrium of a discharge for pulse planning purposes requires to make a prediction on profiles in the core. A natural solution is to couple the inverse equilibrium solver with the predict-first simulations of a transport solver, as discussed above. This strategy is defined here as KEP for Kinetic Equilibrium Prediction (or Preparation). In this paper, we describe the coupling between FBT and RAPTOR, with a similar method to the one presented in \cite{carpanese_first_2020} for the real-time RAPTOR-LIUQE KER. It will be shown that providing the predicted internal profiles to FBT enhances the calculation of the feedforward poloidal field (PF) coil currents (Fig. \ref{fig:legend_coils_polview}) required to maintain the magnetic topology, while the plasma evolves, reducing the risk of disruptions and plasma shape misalignment. At the time of writing, similar kinetic-equilibrium coupling \cite{ostuni_core_2022} has been reported between FEEQS \cite{heumann_quasi-static_2015} and METIS \cite{artaud_metis_2018} at JT-60SA, and is currently under development \cite{battaglia_APS_2025, felici_APS_2025} for SPARC with TORAX \cite{citrin_torax_2024} and GSpulse \cite{wai_feedforward_2025}.

The paper is structured as follow. Section \ref{Sec. KEPsuite} presents the KEP equations and methodology. Section \ref{SubSec:RAPTOR} describes the RAPTOR model used for pre-shot simulations in TCV, including a new model for H-mode transition that extends the Martin scaling law \cite{martin_power_2008} to favorable and unfavorable magnetic configurations. The gradient-based model, originally designed to cover transitions from L to H mode, has been extended to include the ion temperature equation and changes in triangularity, accounting for the higher L-mode performances observed in Negative Triangularity (NT) \cite{coda_enhanced_2021,marinoni_brief_2021,thome_overview_2024,mariani_negative_2024,sauter_operation_2025}. Equations of the inverse equilibrium problem are reminded in Section \ref{SubSec:MEQ}, introducing the FBT-RAPTOR coupling in Section \ref{SubSec:FBT-RAPTOR}. The transport and equilibrium calculations are benchmarked on $211$ shots in Section \ref{Sec:PreShotScan}, statistically validating the models with various shapes and scenarios. Section \ref{Sec. TCVResults} finally reports on the integration of the new coupling solution into the TCV shot preparation system. We discuss how this new preparation enhances the stability and accuracy of the equilibria, along with possible directions for improvement of the present approach.

\section{Physics models and codes}
\label{Sec. KEPsuite}
\subsection{RAPTOR}
\label{SubSec:RAPTOR}
RAPTOR is a fast and light transport simulator for the current density ($j_{\parallel}$), temperatures ($T_{e,i}$) and electron density ($n_e$). In other words, it solves for a non-linear system of PDEs of the form\footnote{More detailed formulations of these equations can be found in \cite{felici_real-time_2011,teplukhina_realistic_2018,van_mulders_full-discharge_2023}}: 
    {\footnotesize
\begin{empheq}[]{alignat=2}
  & \sigma_{\parallel} 2\pi\frac{\hat{\rho}\rho_B^2}{V_{\hat{\rho}}'} && \left(\frac{\partial }{\partial t} - \frac{\hat{\rho} \dot{\Phi}_B}{2\Phi_B}\frac{\partial }{\partial \hat{\rho}}\right)\psi   = \underbrace{\frac{g_a}{\mu_0\hat{\rho}} \frac{\partial}{\partial \hat{\rho}} \left ( \frac{g_b}{\hat{\rho}} \frac{\partial \psi}{\partial \hat{\rho}} \right )}_{\propto j_{\parallel, tot}} - (j_{bs}+j_{cd})\nonumber \notag \\ 
  &\frac{3}{2} (V_{\hat{\rho}}')^{-\frac{5}{3}} && \left(\frac{\partial }{\partial t} - \frac{\dot{\Phi}_B}{2\Phi_B}\frac{\partial }{\partial \hat{\rho}}\hat{\rho}\right)[(V_{\hat{\rho}}')^{\frac{5}{3}}n_{e,i}T_{e,i}]  + \frac{1}{V_{\hat{\rho}}'}\frac{\partial}{\partial \hat{\rho}}  \hat{Q}_{e,i}  =  P_{e,i}, \nonumber \notag\\
  & \frac{1}{V_{\hat{\rho}}'} &&\left(\frac{\partial }{\partial t} - \frac{\dot{\Phi}_B}{2\Phi_B}\frac{\partial }{\partial \hat{\rho}}\hat{\rho}\right) [V_{\hat{\rho}}' n_e]  + \frac{1}{V_{\hat{\rho}}'}\frac{\partial}{\partial \hat{\rho}}\hat{\Gamma}_e  =  S_e, \nonumber \notag
\end{empheq}
}

\noindent using the normalised toroidal flux coordinate $\hat{\rho} = \sqrt{\frac{\Phi}{\Phi_B}}$, with  $\psi \: [\weber]$ being the magnetic poloidal flux, defined following the COCOS-11 convention \cite{sauter_tokamak_2013}, $\Phi\: [\weber]$ the enclosed toroidal flux, $\Phi_B$ its value at the plasma boundary, and given a set of flux-surface averaged current drive source $j_{cd} \:[\ampere.\meter^{-2}]$, bootstrap current $j_{bs} \:[\ampere.\meter^{-2}]$, power $P_{e,i}\:[\watt.\meter^{-3}]$ and particle $S_{e}\: [\meter^{-3}.\second^{-1}]$ source profiles. Information on the geometry is embedded in $g_a =\frac{\hat{\rho}\rho_B^2 T^2}{8\pi V'_{\hat{\rho}}\phi_B^2} \: [\meter^{-3}]$, in the dimension-less coefficient $g_b = \langle \frac{1}{R^2}\rangle\langle \frac{(\Grad V)^2}{R^2}\rangle$, in the non-normalized toroidal radius at the plasma boundary $\rho_B = \sqrt{\frac{\Phi_B}{\pi B_0}}\: [\meter]$, and in the volume radial derivative $V_{\hat{\rho}}'= \frac{dV}{d\hat{\rho}}\: [\meter^3]$, usually computed externally from an equilibrium code, where $\langle \cdot \rangle$ denotes flux-surface average, $R$ the device major radius, $T=RB_\phi$ with $B_\phi$ the toroidal magnetic field. The heat and particle fluxes write 
$Q_{e,i}\: [\watt.\meter^{-2}] =\rho_BV_{\hat{\rho}}^{-1}\hat{Q}_{e,i}\: [\watt]$ and $\Gamma _{e,i}\: [\meter^{-2}.s^{-1}] = \rho_BV_{\hat{\rho}}^{-1}\hat{\Gamma}_e \:[s^{-1}]$. These diffusion equations, obtained by flux-surface averaging of Ohm's law and moments of the Fokker-Planck equation \cite{hinton_theory_1976}, are then discretised using a finite element method by projecting the continuous functions $\{\psi(\hat{\rho},t),T_{e,i}(\hat{\rho},t),n_{e}(\hat{\rho},t)\}$ on a base of $n_{sp}$ cubic splines $\Lambda_\alpha$ \cite{felici_real-time_2011}:
{\small \begin{equation}
\begin{bmatrix}
\boldsymbol{\psi}(\hat{\rho},t)\\
\mathbf{T_{e,i}(\hat{\rho},t)}\\
\mathbf{n_e(\hat{\rho},t)}
\end{bmatrix}
= \sum_{\alpha=1}^{n_{sp}} \Lambda_\alpha(\hat{\rho})
\mathbf{x}^\alpha(t)
, \quad
\mathbf{x}^\alpha(t) = 
\begin{bmatrix}
\boldsymbol{\hat{\psi}}^\alpha(t) \\
\mathbf{\hat{T}_{e,i}}^\alpha(t) \\
\mathbf{\hat{n}_{e}}^\alpha(t)
\end{bmatrix}
\end{equation}}
\noindent and defining $\mathbf{x}(t)$ as the state vector, so that the above transport equations can be rewritten in residual form: 
{\small\begin{equation}
    \forall t, \quad \Tilde{f}(\dot{\mathbf{x}}(t),\mathbf{x}(t),\mathbf{u}(t),\mathbf{p}) = 0, \quad \mathbf{u}(t) = 
    \begin{bmatrix}
I_p(t) \\
P_{aux}(t) \\
\vdots
\end{bmatrix}
\end{equation}}
where $\mathbf{u}$ and $\mathbf{p}$ are the actuator and model parameter vectors and $P_{aux}(t)\:[\watt] = \int_V P_{e,i}(\hat{\rho},t)dV$  is the input power integrated over the total plasma volume. For each time step $k$, a loop of Newton iterations is performed by computing the Jacobian matrices ${\mathcal{J}^k_i = \frac{\partial \Tilde{f}_k}{\partial x_i}}$ until $\mathbf{x}_k$ satisfies ${||\Tilde{f}(\dot{\mathbf{x}}_k,\mathbf{x}_k,\mathbf{u}_k,\mathbf{p})||_2 < \epsilon_{tol}}$, i.e. the residual is below a given tolerance $\epsilon_{tol}$, or until they reach the maximum number of iterations imposed for the simulation.

Particle sources are not yet included in the model, such that $S_{e} = 0$. Similarly, the heat flux $Q_{e,i}$, is assumed to be proportional to $n_{e,i}\chi_{e,i}\frac{\partial T_{e,i}}{\partial \hat{\rho}}$, after neglecting convective and pinch terms, including the particle flux $\Gamma _{e,i}$ contribution. The latter appears in the electron density diffusion equation as: $\Gamma _e \sim -D_e\frac{\partial n_e}{\partial\hat{\rho}} + V_e n_e$. The diffusivity coefficient $\chi_{e,i}$ and $V_e/D_e$ are then to be evaluated through one of the fast transport models implemented in RAPTOR, namely the ad-hoc analytical formulation \cite{felici_real-time_2011_thesis}, Bohm-gyroBohm model \cite{erba_validation_1998}, gradient-based model \cite{teplukhina_simulation_2017} (all based on empirical observations of the plasma confinement) and the QuaLiKiz neural network surrogate model \cite{citrin_real-time_2015}. These reduced models allow RAPTOR to bypass the numerically demanding calculation of tokamak turbulence while retaining the main nonlinear interactions, making it particularly suitable for applications requiring rapid transport simulations. 

In this work, we refer to a \textit{pre-shot simulation} the rapid prediction of a plasma discharge trajectory, given a fixed equilibrium, derived from a set of programmed engineering parameters and validated only against past experiments, to help in the preparation, validation and optimization of a scenario. We call \textit{pulse schedule} the set of engineering parameters and pre-computed information that is required by RAPTOR and FBT before the shot, currently including: requirements for the total plasma current, in-vessel position and shape, engineering parameters of the heating systems (power, absorption factor, etc.) and gas reference. TCV is equipped with two NBI launchers and multiple gyrotrons. We focus in this work on ohmic and NBI heated cases, since NBI is the primary heating system used for H-mode studies in TCV and we focus on L and H modes and their transition. The primary challenge of rapid full discharge simulations lies in the interdependence of transport equations and equilibrium, heating and current drive sources. We present here the choices that have been made in the spirit of keeping the model fast, flexible and easily expandable.

\paragraph{General assumptions.} 
\begin{itemize}
	\item TCV plasmas are mainly composed of (D) deuterium main ion and (C) carbon impurities. Their concentration is determined using the quasi-neutrality equation: 
{\small\begin{equation}
 \label{eq:qzne_manual}
      \frac{n_D}{n_e} = \frac{Z_C-Z_{eff}}{Z_C-1}  \quad \text{and} \quad
      \frac{n_C}{n_e} = \frac{1}{Z_C}\frac{Z_{eff}-1}{Z_C-1},
\end{equation}}
assuming a flat and constant profile of the effective charge $Z_{eff} = 1.5$.
	\item The NBI deposition profiles are modeled by simple Gaussian profiles centered at radius $\hat{\rho}_{dep}$ and a full 1/e width $w_{dep}$ over the normalized radius. The deposited power is then evaluated from the programmed injected power, by removing $15 \%$ of beam duct losses and applying a factor $55\%$ to account for the total absorption into the thermal plasma, notably removing charge exchange losses  \cite{geiger_fast-ion_2017}. Since NBI heating is predominantly central in TCV, it is modeled using a centered deposition profile of width $0.25$ in $\hat{\rho}$ and zero current drive, the latter being relatively small.
\end{itemize}

\begin{table*}
\centering
    \begin{tabular}{||c c c c c c c c c c||}
    \hline
         Mode& Shape & $\lambda_{Te}$ & $\lambda_{Ti}$ & $\lambda_{ne}$  & $T_e |_{sep}\:[\electronvolt]$ & $T_i |_{sep}\:[\electronvolt]$ & $n_e |_{sep} [10^{19}\meter^{-3}]$ & $H_e^{98(y,2)}$ &  $H^{98(y,2)}$\\
         \hline\hline
         L & PT& 3.2 & 3 & 2   & 20 & 16 & 0.5 & 0.4 & 0.7 \\
         \hline
         H & PT& 2.3 & 2.5 &1   & 100 & 80 & 1 & 0.5 & 1  \\
         \hline
         L & NT& 3 & 3 & 3  & 20 & 16 & 0.5 & 0.5 & 1\\
          \hline
    \end{tabular}
     \vspace{1em}
    \caption{Typical parameters used in the gradient-based model depending on the confinement mode and triangularity (positive, PT, or negative, NT) including: the logarithmic gradients $\lambda_\sigma$ and boundary conditions at the separatrix $T_\sigma |_{sep}$, and the electron $H_e^{98(y,2)}$ and total $H^{98(y,2)}$ confinement factors to the ITER H98(y,2) baseline scaling law \cite{confinement_chapter_1999} used as feedback reference for the PI controller.}
 \label{table:MS_parameters}
\end{table*}

\paragraph{Transport and pedestal.} The gradient-based model \cite{teplukhina_simulation_2017} is used to evaluate the heat and particle transport coefficients, $\chi_{e,i}$, $V_{e,i}$ and $D_{e,i}$. This model is based on the observation that radial transport usually behaves in a critical regime, in which the logarithmic gradients in the intermediate region between a sawtooth inversion $\hat{\rho}_{inv}$ and pedestal $\hat{\rho}_{ped}$ radii remain ``stiff" \cite{ryter_confinement_2001,garbet_physics_2004,doyle_chapter_2007,sauter_non-stiffness_2014}. This critical regime is found to emerge at a threshold value of logarithmic temperature gradient $\nabla T/T$, below which ITG and TEM become marginally stable. It is thus predominant in this core region, where the temperature increases towards the magnetic axis, while rarely observed in the colder edge region, whose gradient remains above the threshold \cite{garbet_profile_2004,sauter_non-stiffness_2014}. This observation allows to divide the electron and ion temperatures ($\sigma=T_{e,i}$) and electron density profiles ($\sigma=n_e$) in three separate regions: 
{\small\begin{equation}
-\frac{d\ln{\sigma}}{d\hat{\rho}} = 
\begin{cases}
      0 &\text{for } 0<\hat{\rho}<\hat{\rho}_{inv}\\
      \lambda_{\sigma} & \text{for } \hat{\rho}_{inv}<\hat{\rho}<\hat{\rho}_{ped}\\
       \frac{1}{\sigma} \mu_{\sigma} &\text{for } \hat{\rho}_{ped}<\hat{\rho}<1,\\
\end{cases} 
\end{equation}}
where the linear peripheral gradient $\mu_{\sigma}$ is a non-stiff parameter of the model corresponding to the pedestal gradient. As noted in \cite{sauter_non-stiffness_2014}, L-modes also have a clear pedestal gradient, hence this model allows to unify L- and H-modes stationary profile representations. Full simulations of ASDEX-Upgrade \cite{teplukhina_simulation_2017,van_mulders_scenario_2023}, TCV \cite{teplukhina_simulation_2017} and DEMO \cite{van_mulders_scenario_2023-1} discharges using the RAPTOR gradient-based model confirmed that the inverse scale-length for the electron temperature $\lambda_{T_e} (\sim a/L_{Te})$ and density $\lambda_{n_e}$ are indeed mostly constant across a stiff radial region, extending from $\hat{\rho}_{inv}\sim 0.2$ (depending on the $q=1$ location) to $\hat{\rho}_{ped}\sim 0.8$, and are essentially independent of plasma parameters such as the plasma current $I_p$, power sources $P_{e,i}$, line-averaged density $n_{el}$ or triangularity $\delta$ (as shown in \cite{sauter_non-stiffness_2014}) ; while the linear peripheral gradients $\mu_{T_e,T_i,n_e}$ exhibits a non-stiff behavior, typically responsible for variations in confinement during the discharge. In this model, the value of the non-stiff electron temperature gradient $\mu_{T_e}$ is modified by a PI controller to match a total confinement factor $H=\tau_E/\tau_E^{\text{scal}}$, scaling the energy confinement time obtained during the discharge, $\tau_E = W/P_{loss}$, with a data-driven scaling law $\tau_E^{\text{scal}}$, with $P_{loss} = P_{oh} + P_{aux} -\frac{dW}{dt}$ the power loss. The total thermal energy $W = W_{tot} = W_e + W_i$ or electron energy $W = W_e$ is used in these computations, depending on the model chosen for the ion temperature, as detailed in the next paragraph. Similarly, $\mu_{n_e}$ is controlled to match the line-averaged density $n_{el}$. This model is thus able to follow any change in confinement and plasma density, if known beforehand, by modifying the height of the pedestals in a dynamic way. Boundary conditions at the separatrix $T_\sigma^{sep}=T_\sigma(\hat{\rho}=1)$ are kept constant in each confinement phase of the discharge, following typical values observed experimentally. Their values and those of the logarithmic gradients $\lambda_\sigma$ are reported in Table \ref{table:MS_parameters}. Such simple model eliminates the need for complex pedestal physics, although it does require a starting hypothesis about the confinement quality. In fact, choosing the gradient-based model for the pre-shot simulations was directly dictated by the need to solve the diffusion equations from the magnetic axis to the last closed flux surface, and the need to rely on a simple, robust transport model, which could match most of the TCV experimental data simulated in this work and easily cover L- and H-mode phases and transitions. In the present work, we compute the energy confinement time of reference $\tau_E^{\text{scal}}$ from the ITER H98(y,2) baseline scaling law \cite{confinement_chapter_1999}:
$$\tau_E^{\text{scal}} = 0.0562I_p^{0.93}B^{0.15}n_{el,19}^{0.41}M^{0.19}R^{1.39}a^{0.58}\kappa^{0.78}P_{in}^{-0.69},$$
with $I_p$ the total plasma current, $R$ the device plasma major radius (at the geometric center), $B$ the vacuum magnetic field at radius $R$, $n_{el,19} [10^{19}m^{-3}]$ the line-averaged density, $M$ the average ion mass,  $a$ the minor radius, $\kappa$ the elongation and $P_{in}$ the input power. 

\begin{figure}[htpb!]
    \includegraphics[width=1\linewidth]{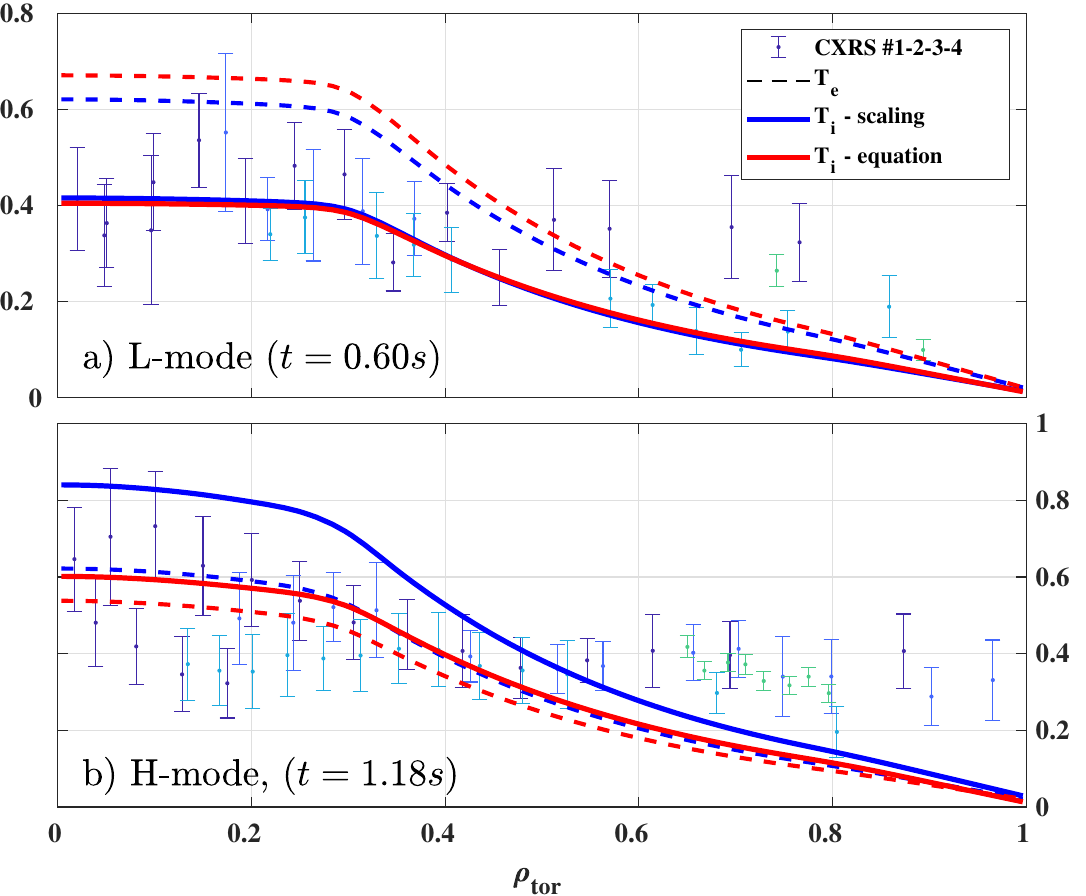}
    \captionsetup{width=\linewidth}
    \vspace{.1em}
    \caption{Comparison of two $T_i$ estimations, using Eq. \ref{equ:PRETOR} with $H_e^{98(y,2)} = 0.5$ (in blue) and solving for the full ion heat equation with $H^{98(y,2)} = 1$ (in red), with the measurements of 4 CXRS systems for shot $\#81882$, and comparison with $T_e$, in the L-mode ohmic phase (a) and in the NBI-heated H-mode phase (b). These models yield similar results, both predicting the CXRS measurements within their error-bars, except in the H-mode outer radii region, where measured values are biased by Edge Localized Modes.}
    \label{fig:81882_ti_profile}
\end{figure}

\paragraph{Ion temperature.} The total confinement factor $H^{98(y,2)}=\tau_E/\tau_E^{98(y,2)}$ is given as reference to the gradient-based model (see Table \ref{table:MS_parameters}) when solving for the ion and electron heat equations together. Contributions from both species to the total transport are determined by the ratio of the ion to electron temperatures at the pedestal top, $\gamma =T_i/T_e |_{ped}$ and the logarithmic gradients $\lambda_{T_{i,e}}$, so that the edge electron temperature gradient is controlled in relation to the total energy content $W_{tot}$. Values at the separatrix are set to $T_i |_{sep} = \gamma T_e |_{ped}$, such that the ion temperature gradient follows $\mu_{T_i} = \gamma \mu_{T_e}$. This fraction $\gamma$ is then defined from the density and NBI power, such that the ion temperature at the pedestal doesn't exceed $90\%$ of the electron temperature in ohmic plasmas and linearly decreases at lower densities, with the ad-hoc expression $\gamma = \min(0.4+0.5n_{e,l}[5.10^{19}m^{-3}],0.9)$, to account for the weaker collisional coupling between ions and electrons at lower densities, while being set at $160\%$ of the electron temperature when NBI heats up the ions. In these phases, NBI heats up both ion and electron species. $60\%$ of the absorbed power is given to the ions in the simulations, while the total power absorbed by the thermal plasma (electrons and ions) is used in $\tau_E^{98(y,2)}$, reducing the need for a precise ion to electron heating ratio. This assumption can be improved in the future by using fast codes like RABBIT \cite{weiland_rabbit_2018} once validated for TCV. In ohmic phases, on the other hand, electrons are heated by the ohmic power and heat up the ions through equi-partition. Solving the ion and electron heat equations together makes it possible to translate the difference in stiff transport between the two species. 

Alternatively, the ion temperature can be scaled from the electron temperature $T_e$, using the ratio: 
{\small\begin{equation}
    T_i/T_e =0.14 \ \frac{n_e}{n_i} Z_{eff}^{-0.5}\left [n_{el} [10^{19} m^{-3}]\right ]^{0.8} q_{95}^{0.6},
    \label{equ:PRETOR}
\end{equation}}
which original formula, commonly used in TCV experimental analysis, was obtained from simulations of ohmic and EC heated scenarios, with $n_i = n_D+n_C$ the total ion density and $q_{95}$ the safety factor at $95\%$ of the enclosed poloidal flux. The dependence of $T_i/T_e$ on the NBI power is implicitly given through the line-averaged density term. In this model, the confinement factor used for controlling the electron pedestal gradient is set to: $H_e^{98(y,2)}=\left ( \frac{W_e}{W_{tot}}\right )H^{98(y,2)}$, with $W_e$ the electron thermal energy. This simple scaling reduces the number of equations to be solved and thus, the simulation time (by $10-15\%$) and the risk of numerical oscillation due to non-linear interactions between ion and electron temperatures, but does not take into account the differences in transport and heat sources between the two species. Both models tend to give reasonable profiles for most of the shots. As can be seen for example in Fig. \ref{fig:81882_ti_profile}, the accuracy of TCV CXRS measurements does not always allow one model to be preferred over another. In particular, predicted temperature profiles are highly influenced by their value at the pedestal top and during ELMy H-mode phases, where CXRS measurements tend to be less accurate. However, statistical comparison over a large shot database, presented in Sec. \ref{Sec:PreShotScan}, show that solving for the ion heat equation gives more correct results than the scaling in Eq. \ref{equ:PRETOR}, at the cost of a greater sensitivity of the model gradients control and thus the need for a higher time resolution. 

\begin{figure}[bp!]
    \centering
    \includegraphics[width=\linewidth]{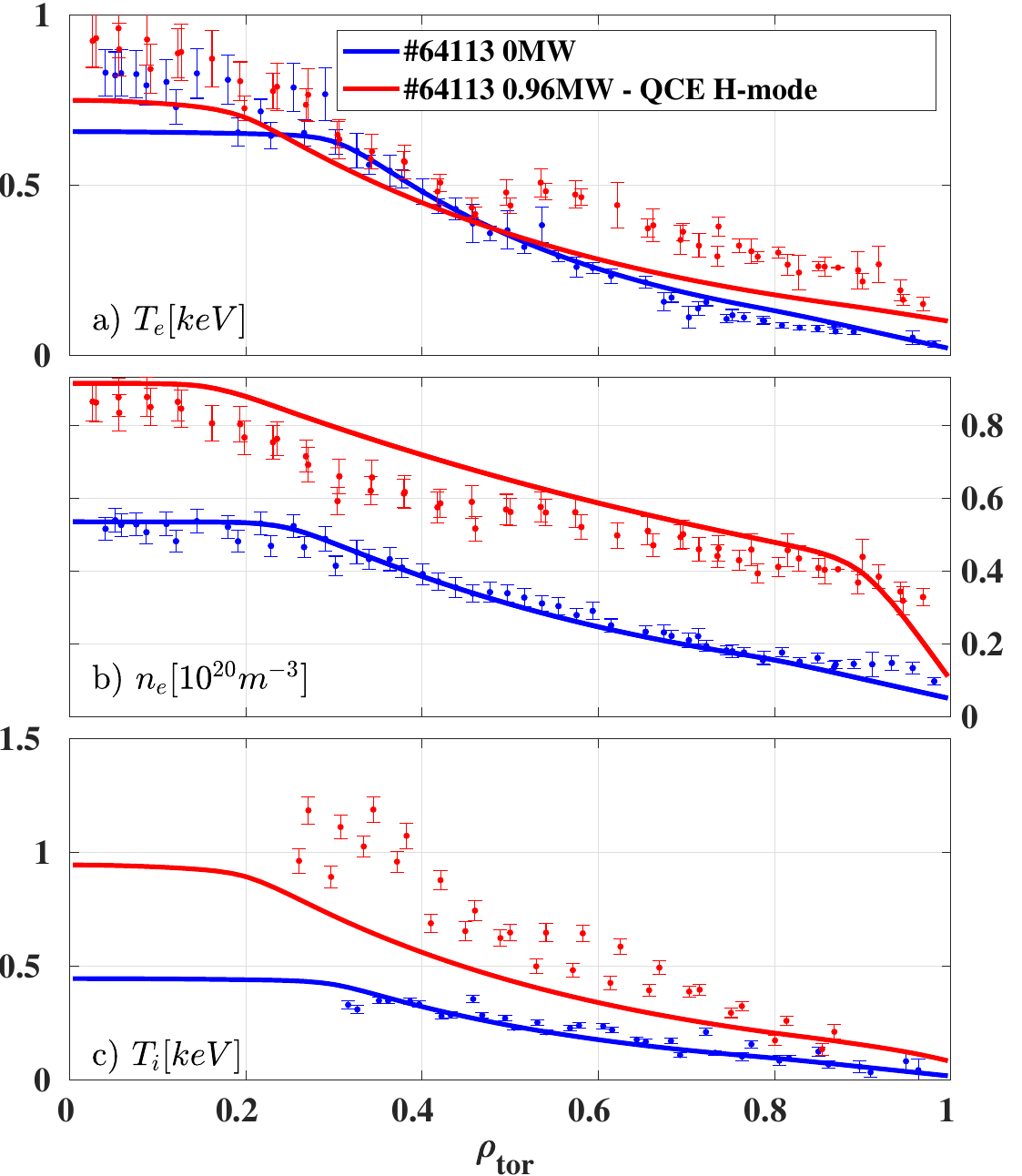}
    \vspace{0.1em}
    \caption{Evolution of (a) $T_e$, (b) $n_e$ and (c) $T_i$ profiles before and after the onset of NBI. As the H-mode is triggered and reaches a QCE regime, the TS measurements (markers with error-bars) show a slightly better confinement ($H_{\text{exp}}^{98(y,2)} > 1$) than the one modeled in RAPTOR (plain line), particularly shown in $T_i$. The experimental $n_{e,l}$ was used as input to this simulation.} 
    \label{fig:PreShotScan:QCE_NBI_profiles}
\end{figure}

\paragraph{L-H transitions. }Transition from low (L) to high (H) confinement mode is triggered during the simulation when the power going through the separatrix verifies:

\begin{equation}
    P_{sep} = P_{loss} - P_{rad} = P_{oh} + P_{aux} - P_{rad} -\frac{dW}{dt}> P_{LH},
    \label{eq:PsepPLH}
\end{equation}

provided that $n_{el}$ is above a minimum threshold, of around $3.10^{19} \meter^{-3}$ for TCV. As in many carbon devices, the core radiated power can be neglected \cite{labit_lh_2025} such that we can assume $P_{sep} = P_{loss}$. Since the $\frac{dW}{dt}$ term is highly sensitive to variations in the controlled $\mu$ gradients, yet remains very small over most of the discharge, it is for now neglected. Following the same strategy as in \cite{van_mulders_rapid_2021}, the power threshold $P_{LH}$ follows the 2008 Martin scaling law \cite{martin_power_2008}: 
{\small\begin{equation}
    P_{LH} = \frac{2}{A_i}2.15\ 10^6 \left [n_{el} [10^{20} m^{-3}]\right ]^{0.782 }B_0^{0.772}a^{0.975}R_0^{0.999} 
\end{equation}}
This scaling was obtained from a multi-machine database of single null diverted configurations with ion grad B drift towards the X-point (Fwd $B_t$), $q_{95} > 2.5$ and neglecting cases with high radiation losses ($P_{rad}/P_{loss} < 50\%$). It notably excludes limited cases, where the H-mode access is rarely observed, but also double-null (DN) and SN configurations with unfavorable ion grad B drift (Rev $B_t$). Indeed, it has been shown \cite{hinton_neoclassical_1985,wagner_experimental_1985}  that the H-mode access in SN diverted geometries strongly depends on the direction of the ion grad B drift, $\Vec{v}_{D,i} = \frac{\mu}{2q_i B^2}\B\times \Grad B$, and is systematically facilitated when the drift points \textit{towards} the X-point location. On the other hand, the threshold for DN geometries remains uncertain. While an increase of $P_{LH}$ has been reported in various machines \cite{asdex-team_h-mode_1989,wagner_experimental_1985,meyer_formation_2004}, experiments on MAST-U \cite{meyer_formation_2004}, later reproduced on ASDEX Upgrade and NSTX \cite{meyer_h-mode_2005}, suggest that a lower threshold can be found in a true double-null configuration, i.e. provided that the two flux surfaces passing by the X-point are almost identical. This connected DN (CDN) can be characterized by a near-zero radial separation of the two X-point flux surfaces at the LFS mid-plane, $\delta r_{sep} < \delta r_{\text{thr}}$, with $2\delta r_{\text{thr}}$ being of the order of the ion thermal Larmor radius $\hat{\rho}_{i}$ or the SOL decay length $\lambda_{SOL}$, with SOL the scrape-off layer. This criterion can be continuously evaluated from one configuration to another, for both SN and DN configurations, choosing $\delta r_{sep}$ equals to infinity or to the radial gap between the Last Closed Flux Surface (LCFS) and the wall when the secondary X-point is outside of the vacuum vessel (or limiter contour), and can be easily included in the prediction using the FBT fluxes and mapping. However, controlling precisely $\delta r_{sep}$ is difficult to achieve experimentally, hence model predictions should be treated cautiously. 
\label{Par:lh-transition}

\begin{figure}[htpb!]
    \includegraphics[width=\linewidth]{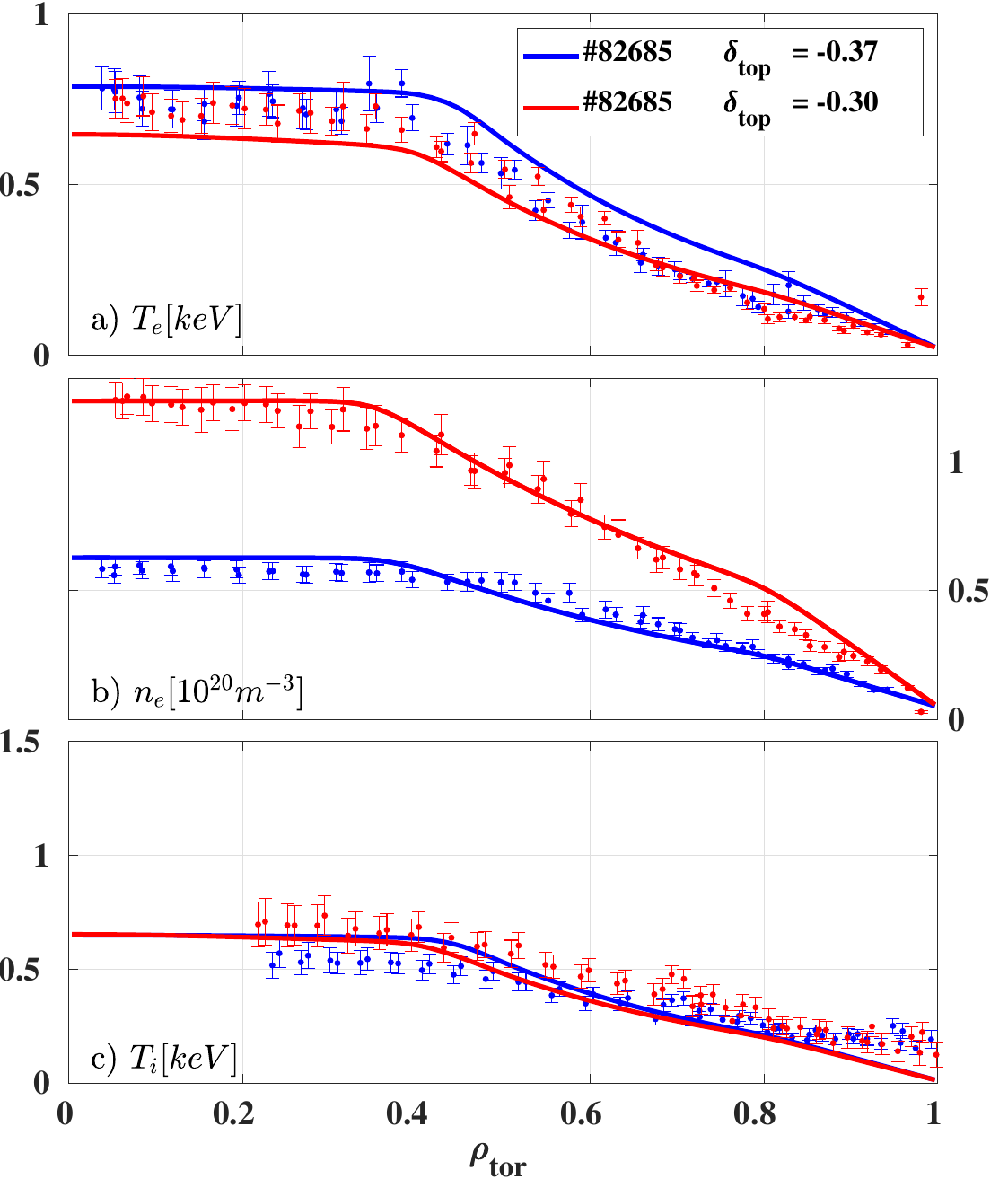}
    \captionsetup{width=\linewidth}
    \vspace{0.1em}
    \caption{(a) $T_e$, (b) $n_e$ and (c) $T_i$ profiles of a negative triangularity (NT) shot. As the shaping decreases from $\delta = -0.37$ to $-0.30$, the NBI power is switched on and ramped up to $516kW$, leading to an increase in density and total energy. With an $H^{98(y,2)}=1$ and taking the experimental $n_{e,l}$, RAPTOR (plain line) matches the high confinement properties of the L-mode NT plasma, slightly over-estimating $T_e$ in the ohmic (lower density) phase. }
    \label{fig:Te_ne_ti_profiles_ohmic_NT_82685}
\end{figure}

To include these various remarks, the power threshold $P_{LH}$ is re-scaled with an ad-hoc factor $\alpha_{LH}$, whose value is determined from the divertor geometry programmed in FBT, as:
{\small\begin{equation}
\alpha_{LH} = 
\begin{cases}
      \alpha_l, &\textit{(limited)}\\
      f_{\text{div}}(s) =  \alpha_{f}+(\alpha_{u}-\alpha_{f})G(s)  , & \textit{(diverted)}\\
\end{cases} 
\label{eq:alphaLH}
\end{equation}}

\noindent where $s = \sigma_{fav}|\frac{\delta r_{sep}}{\delta r_{\text{thr}}}|$, $\sigma_{fav}=\pm 1$ is a factor which is positive in favorable configuration and negative in unfavorable configuration, $G(s)=\frac{ 1}{1+e^{s}}$. Thus, $f_{\text{div}}(s)$ is a continuous function of $s$ (depicted in Fig. \ref{fig:fiv_curve} below) that is characterized by two horizontal asymptotes, $\underset{+\infty}{\lim}f_{\text{div}}(s) = \alpha_{f} = 1$ (favorable, Fwd $B_t$)) and $\underset{-\infty}{\lim}f_{\text{div}}(s) = \alpha_{u} = 2$ (unfavorable, Rev $B_t$), with ${\lim}f_{\text{div}}(0)=\frac{\alpha_f+\alpha_u}{2}=1.5$ for a perfectly connected DN. In this work, $\alpha_l$ is set to a high prohibitive value,  but could later be refined to include particular cases of H-mode access in limited configuration. In practice, most SN discharges that receive the power required to transition to H-mode have a clear favorable configuration such that \textit{s} takes on large positive values. This is the case, for example, in the shot $\#64113$ (Fig. \ref{fig:PreShotScan:QCE_NBI_profiles}), where the equilibrium configuration and NBI power were programmed with the aim to obtain an H-mode in Quasi-Continuous Exhaust (QCE) regime \cite{FAITSCH2021100890}. If CDN configurations are later found to be favorable to L-H transitions, this strategy can be adapted by replacing $f_\text{div}$ by a function of the form: 
{\small\begin{equation}
      f_{\text{div}}^*(s) = G(-s)[\alpha_{f}-\eta e^{-\frac{s^2}{2}}]+G(s)[\alpha_{u}-(\alpha_{u}-\alpha_{f}+\eta) e^{-\frac{s^2}{2}}]
\label{eq:alphaLH_star}
\end{equation}}
\noindent such that in perfectly connected DN, $\alpha_{LH} = f_{\text{div}}^*(0) = \alpha_f(1-\eta)$ is lower than 1. It should be noted, however, that the quality of this prediction remains determined by our ability to control precisely the value of $\delta r_{sep}$ during the discharge, the discrepancy between the equilibrium programmed in FBT and that obtained experimentally being outside the scope of this work.

\paragraph{Extension of the gradient-based model to negative triangularity plasmas.} NT plasmas are less prone to transition to H-mode than standard PT plasmas \cite{nelson_prospects_2022,nelson_robust_2023}, while often achieving better L-mode performances. Recent results \cite{coda_enhanced_2021} showed that, in TCV, NT L-mode plasmas heated by NBI achieve a total pressure comparable to that observed in PT H-mode, while not switching to H-mode, even with a favorable X-point configuration, with sufficiently negative minimum triangularity \cite{Sauter_2023}. In this work, values of $\alpha_{LH}$ listed in Eq. \ref{eq:alphaLH} are increased by a prohibitive factor, such that H-mode access is inhibited in NT. Exact values can later be determined by modeling the access to the second ballooning stability, for example with CHEASE using the ballooning option \cite{lutjens_chease_1996} or coupled to the BALOO code \cite{nelson_prospects_2022,wilson_characterizing_2024}.

A new type of confinement state and associated peaking factors are added to the gradient-based model, which can now transition from positive to negative triangularity transport the same way as from L- to H- mode. In this work, plasma shape is considered NT when $\min (\delta_{up},\delta_{bot}) < -0.2$, with $\delta_{up}$ and $\delta_{bot}$ the upper and bottom triangularities. A smooth transition of $100 \milli \second$ is given to the controller to adapt the confinement properties, starting from the start of the NT shape, then ramping linearly from PT to NT parametrization. Fig. \ref{fig:Te_ne_ti_profiles_ohmic_NT_82685} shows an example of predictive simulation for a JET-like shaped NT plasma with an NBI power ramp, using NT parameters listed in Table \ref{table:MS_parameters}. These parameters, selected to match the set of discharges simulated in this study  (41 NT shots, among the 211 shots presented in Section \ref{Sec:PreShotScan}), are kept the same and constant across the whole database simulated in this work. 

\paragraph{Line-averaged density.} Governed by core-edge interactions, gas valve opening and auxiliary heating, the line-averaged electron density required for the gradient-based model can be complex to predict before the shot. While in many cases this can be reused from a previous experiment, it is also common for the density programming to be modified, making modeling a necessary step of the KEP. In TCV, the gas flow is, by default, feedback controlled on the FIR central chord measurements. In this case $n_{el}$ can be modeled by a simple conversion from the FIR central chord reference to the line-averaged density. However, a strategy was required for phases in which feedback control is deactivated (feedforward only) or density is more difficult to control, such as NBI heated and H-mode phases. For these cases we therefore introduce a set of ad-hoc rules built on the shot programming, which are based on empirical observations of many discharges:
\begin{itemize}
    \item If no heating power is expected and the feedforward gas input is zero, $n_{el}$ undergoes an exponential decay with an ad-hoc lifetime of $125\milli\second$, reaching half its value, and is increased again if the plasma enters the limited regime
    \item if NBI is on, $n_{el}$ is increased by $30\%$ to account for the neutral beam fueling and higher wall recycling.
\end{itemize}
The final ad-hoc trajectory is then increased by $30\%$ when an H-mode is detected within the simulation, accounting for the better confinement and stronger wall recycling. These assumptions are TCV-related, but allow to predict a large variety of cases, from feedback to feedforward density control, and L- to H-modes, with simple conditions easy to test a posteriori. As discussed later, a proper generic density predictive model is needed to move forward, but is outside the scope of this paper.
Meanwhile, we compare four different predictive simulations, on the one hand using either $n_{el}$ from the experiment (E) or predicted from the pulse schedule and the model described above (P); and on the other hand solving for the ion temperature (1) or using a scaling factor from $T_e$ and Eq. \ref{equ:PRETOR}: E1, E2, P1 or P2 as summarized in Table \ref{table:EP_models_summary}.

\begin{table}[ht!]
    \centering
    \begin{tabularx}{\linewidth} {|c|X||c|X|}
    \hline
         \textbf{E} & {experimental \small $n_{e,l}$} & \textbf{1} & {\small $T_i$ solved}\\
     \hline
         \textbf{P} & {\small pre-shot $n_{e,l}$} & \textbf{2} & {\small$T_i/T_e$ from Eq. \ref{equ:PRETOR}} \\
    \hline
    \end{tabularx}
    \vspace{1em}
    \caption{Four model choices used in the KEP simulations for the electron density (E or P) and ion temperature (1 or 2) time evolutions.}
    \label{table:EP_models_summary}
\end{table}

\paragraph{Runtime optimization.} The duration of a RAPTOR simulation is a decisive factor in making the KEP effective for tokamak operations. A simulation shorter than the latency between two discharges (typically between $5$ and $15$ minutes at TCV) allows to quickly validate and use the result of a new pulse schedule. Indeed, while the preparation of experiments can afford longer simulation times, the objective of the present KEP is to enable the rapid testing of various assumptions and to prepare the equilibrium of an experiment, without slowing down operations. Three factors mostly contribute to the computational time of the present KEP: the number of time steps used in RAPTOR, the evaluation of transport in the gradient-based model and the simultaneous resolution of the poloidal flux evolution and three kinetic profiles, $T_e$, $n_e$ and $T_i$. Three strategies are used to reduce the computational time. 

First, a fixed but non-equidistant time step is chosen using information from the pulse schedule, allowing for a higher time resolution in transients and a lower one in stationary phases. Transition times are defined by a criterion on the time derivative of the programmed plasma current $\dot{I}_{p}$, auxiliary power $\dot{P}_{e,i}$, elongation $\dot{\kappa}$ and triangularity $\dot{\delta}$. Keeping in mind that the model controller performs better if the time step is about an order of magnitude lower than the energy confinement time, $\tau_E$, which ranges from milliseconds to tens of milliseconds in TCV, a minimum value $dt_{min}\leq0.5\milli\second$ is chosen for the sharp transitions, and a large $dt_{max}$ up to $1\milli\second$ is accepted elsewhere, with a smooth transition ensuring numerical continuity. This approach, driven by the knowledge of transient phases where shorter physical time scales are expected, enables to find a compromise between the simulation time and the model's ability to capture sudden changes in plasma dynamics. A natural extension is to introduce a robust adaptive time stepping scheme. Second, a fixed scaling for the bulk $n_i$ and carbon $n_c$ ion densities is imposed, as shown in Eq. \ref{eq:qzne_manual}, avoiding extra quasi-neutrality matrix calculations. Finally, the Matlab solver body is generated and compiled in C, reducing the total simulation time by a factor of $2$. With about $3$ to $5$ Newton iterations per time steps, the total runtime, on a standard non-dedicated CPU\footnote{Simulations were run in MATLAB R2017a (Parallel Computing Toolbox) using four concurrent workers on a shared compute server (2 × 8-core Intel Xeon E5-2660, 128 GB RAM; Fedora 22, x86\_64). Reported runtimes are wall-clock times and may vary due to server load and MATLAB internal multithreading (up to 16 threads per worker)}, computed over simulations of $211$ experiments (cf. Sec. \ref{Sec:PreShotScan}) and solving for the ion heat equation finally scales down to $2.3\pm0.38 \minute$ per second of discharge and $12.35\pm 7.36\second$ per average confinement time\footnote{This is an average, over all simulations, of runtime values normalized by the confinement time $\tau_E=W/P_{loss}$ averaged over the total duration of each simulation. The confinement time varies with various plasma parameters, e.g. its size and confinement regime. A lower confinement time requires a higher time resolution, hence increasing the runtime per duration of discharge.}  $\tau_E$, while still leaving room for further optimization.

\subsection{FBT}
\label{SubSec:MEQ}

The MHD equilibrium of a toroidally confined plasma is a solution $\psi(R,Z)$ of the Grad-Shafranov equation \cite{grad_proceedings_1958,shafranov_magnetohydrodynamical_1958}, which describes the force balance $\Grad p = \Vec{j}\times\Vec{B}$ as a function of the pressure profile $p(\psi(R,Z))$ and flux function $T(\psi(R,Z)) = RB_{\phi}$ for an axisymmetric plasma. This solution is essential for solving any type of radial transport problem, as it determines the geometry at any radius. In RAPTOR, this must be provided externally by an equilibrium solver, whose choice depends on the nature of the problem. Among the available options, FBT \cite{hofmann_fbt_1988} and LIUQE \cite{moret_tokamak_2015} from the MEQ suite are both categorized as \textit{free boundary}, as they include in their computation the shape and location of the Last Closed Flux Surface (LCFS) as a function of the currents flowing through the external conductors. However, their scope differs: FBT is used before the shot to determine the coil currents required to maintain a certain shape (\textit{inverse problem}), while LIUQE uses magnetic measurements to reconstruct the effective shape after or in real-time during the discharge (\textit{reconstruction problem}). 

\paragraph{Coil currents optimization.} FBT is used in static mode: impacts of the plasma dynamics and passive vessel currents are neglected, such that the combination of active currents $\boldsymbol{I_a}$ is estimated by minimizing the cost function: 

{\footnotesize
\begin{equation}
    \chi^2 = 
   \underbrace{\sum_{\mu=1}^{N_{C}}w_\mu^C \delta \psi_\mu^2}
    _{ \shortstack{{\scriptsize poloidal flux}\\{\scriptsize error} } } 
   + \underbrace{\sum_{j=1}^{N_{a}} w^P_j R_j  I_j^2}
    _{ \shortstack{{\scriptsize total resistive}\\{\scriptsize power}} } 
    + \underbrace{\sum_{j=1}^{N_{a}} w^T_j(I_j-I_{j-1})^2}
    _{ \shortstack{{\scriptsize current diff. in}\\{\scriptsize adjacent coils}} } + (...),
    \label{Eq:FBT_optimization}
\end{equation}}

\noindent with $(...)$ denoting a range of additional optimization terms, including extra constrains on the magnetic geometry. Following the notation in \cite{carpanese_development_2021}, we call $\boldsymbol{I_a}$ and $\boldsymbol{I_y}$ the sets of filamentary currents flowing through the active coils and plasma, respectively, on the computational grid $X$, such that $\boldsymbol{I_y}=\{j_{pl}(R_{j},Z_{j})\}$ with ($R_{j},Z_{j}$) a point of the plasma domain (inside the LCFS) and $j_{pl}$ the toroidal plasma current density. The poloidal flux error $\delta \psi_\mu^2$ in \ref{Eq:FBT_optimization} is expressed as a distance of the solution $\psi$ at a set of $N_c$ control points ($R_{\mu},Z_{\mu}$) to an optimization parameter $F_b$:
\begin{equation}
    \delta \psi_\mu^2 = ||\psi(R_\mu,Z_\mu,\boldsymbol{I_a},p',TT') - F_b ||^2.
    \label{eq:delta_psi_FBT}
\end{equation}
The ($R_{\mu},Z_{\mu}$) control points do not necessarily end up on the LCFS in the optimization result, but are most often chosen as such. It is common for example, when scheduling a diverted equilibrium, to impose the presence of a magnetic null in the vicinity of these points, where the primary X-point is expected, by adding extra constrains in $\chi^2$. $F_b$ is not known a priori but is a free optimization parameter introduced to minimize flux differences among the control points. The poloidal magnetic flux in \ref{eq:delta_psi_FBT} is computed from the Poisson like equation: 
{\footnotesize
\begin{empheq}[]{alignat=2}
& \Delta^* \psi && = -2\pi R\mu_0  j_{pl}\label{eq:Poisson}\\ 
&  && = \begin{cases}
    -4\pi^2 \left(\mu_0 R^2 p'(\psi) + TT'(\psi) \right) & \text{in } \Omega_{\text{pl}} \label{eq:Grad-Shafranov}\\ 
    0  & \text{in } \Omega_{\text{vac}}, 
    \end{cases}  
\end{empheq}
}

\noindent where $\Omega_{\text{pl}}$ and $\Omega_{\text{vac}}$ are the plasma and vacuum domains respectively, and eq. \ref{eq:Grad-Shafranov} is the Grad-Shafranov equation (GSE), function of the internal plasma profiles $\{p'=\frac{dp}{d\psi}, TT'=\frac{1}{2}\frac{dT^2}{d\psi}\}$. Note that FBT uses the general MEQ convention for coordinates, COCOS=17 \cite{Sauter_2013}. Finally, the value of the poloidal flux at the computational boundary $\delta \Omega_{\text{c}}$, results from a sum of contributions from the plasma and active coil currents: 
\begin{equation} 
    \boldsymbol{\psi_b} =\mathbb{M}_{by} \boldsymbol{I_y}  +   \mathbb{M}_{ba} \boldsymbol{I_a} \qquad\text{on } \delta \Omega_{\text{c}}
    \label{eq:plasma2coils_contribution}
\end{equation}
which define the boundary conditions for eq. \ref{eq:Poisson}
\begin{figure}[htp!]
    \includegraphics[width=\linewidth]{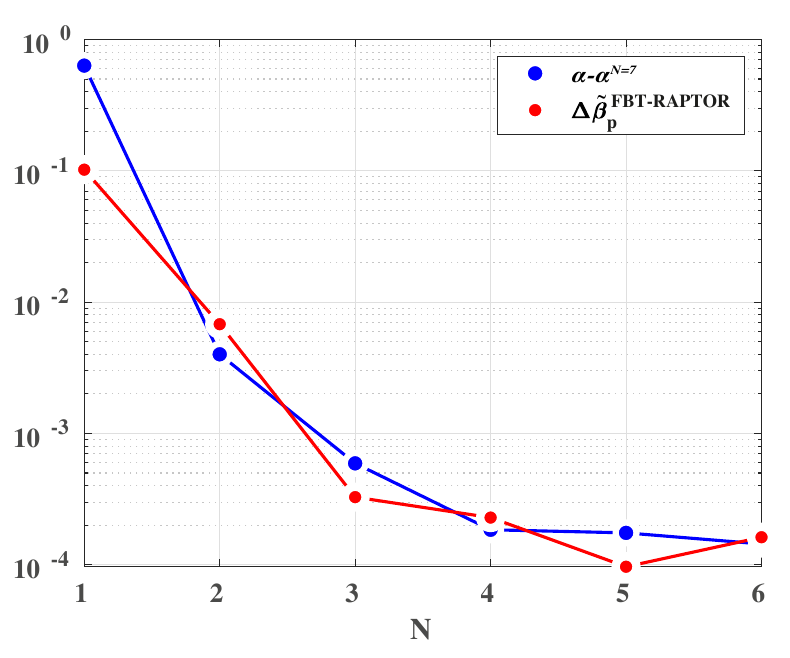}
    \captionsetup{width=\linewidth}
    \caption{Result of the FBT-RAPTOR convergence for TCV shot \#81882 after N iterations. $\alpha=p'_{\text{FBT}}/p'_{\text{RAPTOR}}$ and $\Delta\beta_{pol}=|\beta_{pol}^{\text{FBT}}-\beta_{pol}^{\text{RAPTOR}}|$. The y-axis shows the distance between these quantities and their final values at N=7, with $\Delta{\tilde{\beta}_{pol}}=\Delta\beta_{pol}-\Delta\beta_{pol}^{N=7}$.}
    \label{fig:convergence-81882}
\end{figure}

\subsection{FBT-RAPTOR coupling}

\label{SubSec:FBT-RAPTOR}
\label{Par:FBT-RAPTORcoupling}
\begin{figure*}[ht!]
\centering
\includegraphics[width=0.8\linewidth]{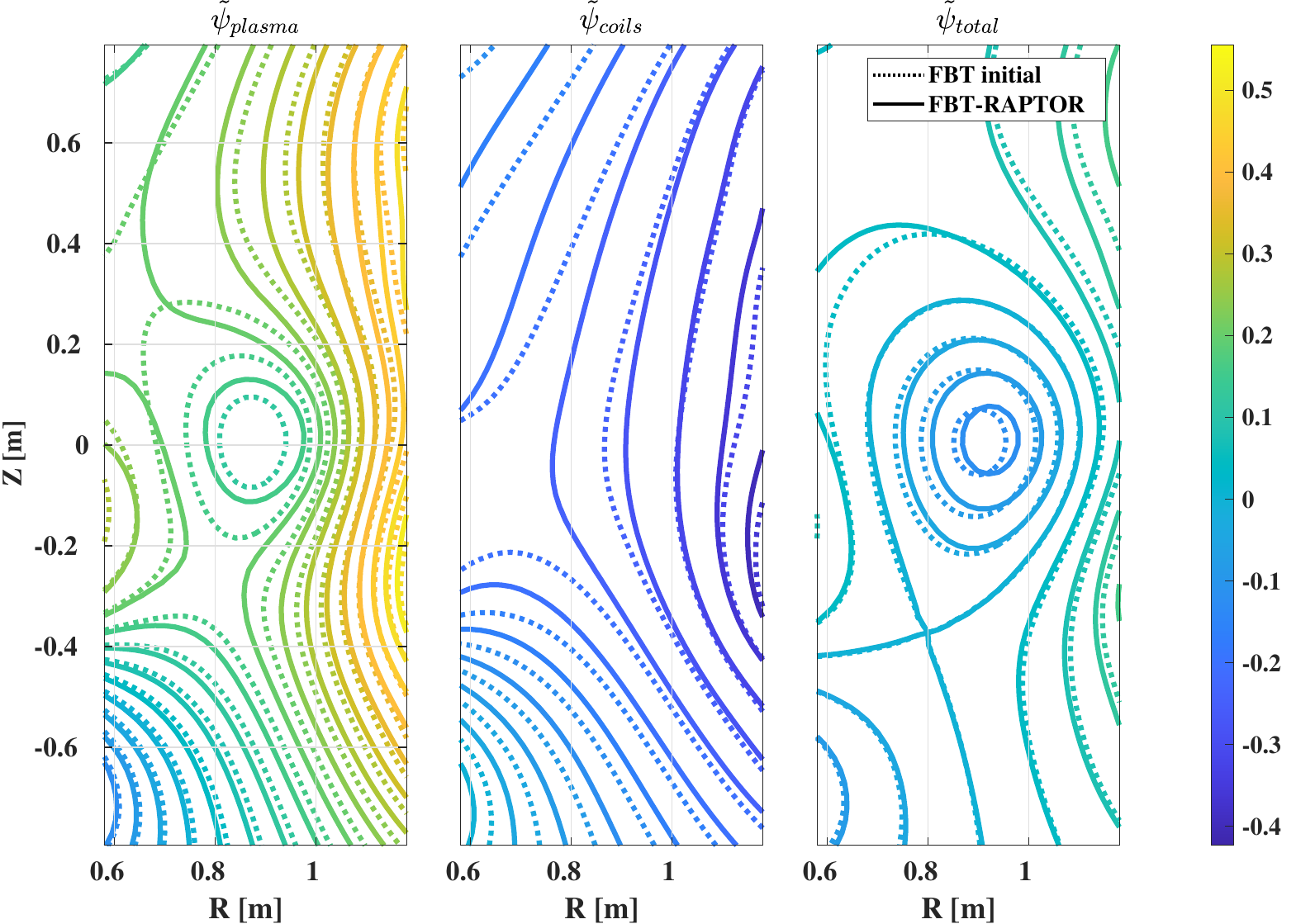}
\vspace{0.5em}
\caption{Plasma and coils contribution to the total poloidal flux in the H-mode phase ($t=1.3\second$) of \#81882 centered around the total flux at the plasma boundary, $\tilde{\psi}=\psi-\psi_{B}$. Due to differences in poloidal beta, $\beta_{pol}^{\text{initial}}<\beta_{pol}^{\text{FBT-RAPTOR}}$, and internal inductance, $l_{i3}^{\text{initial}}<l_{i3}^{\text{FBT-RAPTOR}}$, the coupling (solid line) accounts for a stronger Shafranov shift $\Delta(\hat{\rho})$ than the initial FBT calculation (in dotted line)}
\label{fig:plasma_flux_comparison_initial_vs_profile}
\end{figure*}
With no prior estimation of the kinetic evolution and non-inductive current drive, the right-hand side of the GSE remains under-determined. In this case, the two free functions $p'(\psi)$ and $TT'(\psi)$ are usually built on simple basis functions which take, for the majority of TCV shots, the polynomial form\footnote{General formulation of these standards basis functions accept any positive real values of powers of $\hat{\psi}$ (including non-integer values) and is therefore not strictly speaking polynomial.}: 
\begin{alignat}{1}
    & p'|_\mathrm{FBT}(\psi) = \frac{1}{2\pi}a_{g=1}^{p}(1-\hat{\psi}), \\
    & TT'|_\mathrm{FBT}(\psi) = \frac{\mu_0}{2\pi }\sum^{N_{T}=2}_{g=1}a_g^{T}(1-\hat{\psi})(-\hat{\psi})^{g-1},
    \label{eq:pprime_ttprime_polynoms}
\end{alignat}

with $\hat{\psi}(R,Z)=(\psi(R,Z)-\psi_A)/(\psi_B-\psi_A)$ the normalised poloidal flux, $\psi_A$ and $\psi_B$ the values of the poloidal flux at the plasma axis and boundary; and it is standard practice for operators to parametrize their $a_g$ coefficients (namely, to scale the plasma pressure and on-axis $q$) based on their experience of a given scenario. This method has remained in use for many years at TCV, thanks to the teams' expertise and to the relative robustness of the shape to differences in internal profiles. However:
\begin{enumerate}
    \item its immediate success depends on operators experience, acquired through many years of practice 
    \item  changes in parameters are sometimes driven by operational needs rather than by a realistic predictive estimate (e.g. it is common practice to assume a flatter current distribution (higher $q_A$) in FBT, making sure the experimental inductance is higher or equal to the one assumed in FBT, with the aim of lowering elongation experimentally to approach it from below.)
    \item transition times, e.g. from L to H mode, can be difficult to anticipate accurately without a model
    \item regardless of the quality of the programming, the simple formulation of $p'(\psi)$ and $TT'(\psi)$, typically (eq. \ref{eq:pprime_ttprime_polynoms}) linear for the former and quadratic for the latter, cannot be a correct description of actual plasma profiles, especially in H-modes.
\end{enumerate}

These approximations affect the plasma current in the Poisson like (Grad-Shafranov) equation (\ref{eq:Grad-Shafranov}) and its boundary condition (\ref{eq:plasma2coils_contribution}), hence affecting the poloidal flux error and result of the PF coil currents optimization. 
First of all, the choice of $\beta_{pol}$ and $q_A$ (or $l_i$) largely determines the Shafranov shift $\Delta(\hat{\rho})$, i.e. the outward radial displacement of the magnetic surfaces with respect to the geometric (cylindrical) center of the plasma. This shift comes with an increased compression of magnetic surfaces in the Low Field Side (LFS) of the core plasma, inside the separatrix, and is due to radial forces that tend to drive the whole plasma outward. If $\beta_{pol}+l_i/2$ is underestimated in the calculation, the vertical magnetic field virtually applied by the PF coils to compensate for these forces \cite{mukhovatov_plasma_1971} at the separatrix is underestimated, with the risk of introducing errors in the radial plasma position.
Moreover, as stated in $2$, the distribution of current in the plasma column influences the strength of the quadrupole term added to the vertical field to maintain the vertical plasma stability with a certain elongation \cite{mukhovatov_plasma_1971}. Quantities impacting the growth rate of a vertical instability --- for example, the plasma radial position, elongation and inner and outer gaps \cite{marchioni_vertical_2024} --- are therefore not immune to inaccuracies in $p'$ and $TT'$ profiles, whose consequences may range from a slight misalignment of the shape to a vertical displacement event (VDE). To address these limitations, we propose to constrain $p'(\psi)$ and $TT'(\psi)$ using the RAPTOR pre-shot model presented in Sec. \ref{SubSec:RAPTOR}. In this section and for the rest of the paper, we express the poloidal beta and internal inductance by their ITER design definition \cite{jackson_iter_2008}, used in both FBT and RAPTOR: 
\begin{equation}
    \beta_{p} = \frac{8}{3}\frac{ W_{th}}{\mu_0 R_0 I_P^2} \equiv \frac{2}{3}\frac{W_{th}}{W_{pol}}, \quad l_{i3} = \frac{2V\langle B_{pol}^2 \rangle_V}{\mu_0^2I_p^2R_0}.
\end{equation}
where $W_{th}$ is the total thermal energy, $B_{pol}$ the poloidal magnetic field, $\langle B_{pol}^2 \rangle_V=\frac{1}{V}\int B_p^2dV$ and $V$ is the plasma volume. 

\begin{figure}[htp!]
    \centering
    \includegraphics[width=0.95\linewidth]{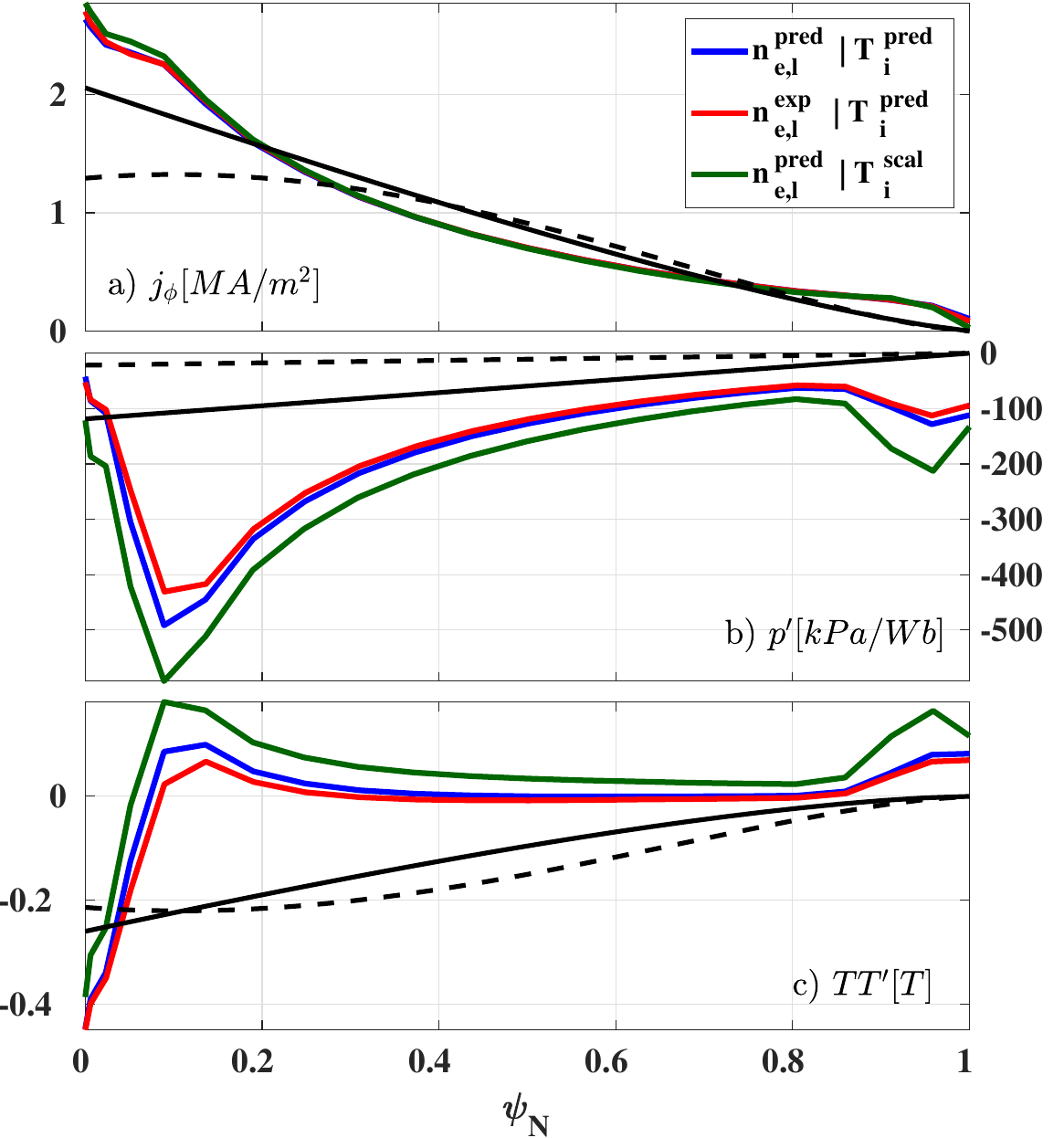}
    \vspace{0.5em}
    \caption{Comparison of RAPTOR $j_{\Phi}$ (a), $p'$ (b) and $TT'$ (c) profiles obtained at $t=2\second$ with models, Table \ref{table:EP_models_summary}, P1 (blue), P2 (green) and E1 (red), i.e. using the pre-shot line-averaged density (blue, green), which is overestimated by about $20\%$ at that time, or experimental post-shot values (from shot $\#83738$, red), and using the predicted $T_i$ (blue, red) or the electron-to-ion scaling (green); in black are the polynomial FBT profiles: standard set-up (solid line) and with low $\beta_{pol}$ and low $l_{i3}$ (dashed line).}
    \label{fig:83738:jtor_pprime_ttprime_comparison}
\end{figure}

The $p'$ and $TT'$ profiles predicted in RAPTOR are interpolated on the equilibrium time grid and given as basis functions in FBT with scaling coefficients $\alpha$ and $\beta$:
\begin{alignat}{2}
    & p'|_\mathrm{FBT}(\psi) &&= \alpha \: p'|_\mathrm{RAPTOR}(\psi), \\
    & TT'|_\mathrm{FBT}(\psi) &&= \beta \: TT'|_\mathrm{RAPTOR}(\psi).
    \label{eq:pprime_ttprime_new_basis}
\end{alignat}
We choose to keep one degree of freedom from the RAPTOR profiles: a small $\Delta\alpha=|\alpha-1|$ correction is applied in FBT to follow the same plasma current target $I_p$ despite small numerical discrepancies, while strictly keeping $TT'$ from the RAPTOR simulation ($\beta=1$). The previous solution for the plasma current distribution, $\boldsymbol{I_y}$, is used as an initial guess to the new FBT run to facilitate convergence. If an equilibrium solution is not reached after 20 iterations, the plasma domain $\partial\Omega_{pl}$ is frozen in the FBT optimization procedure, which resumes until convergence. This approach has proven effective, as non-convergence typically results from small oscillations between two points on the computational grid, and allows a solution to be found for nearly all the equilibria, while keeping the (relatively coarse) standard mesh used in TCV, $n_R\times n_Z=28\times65$. When an iteration of the coupling does not converge at a given time, the previous converged solution for that time is taken. Thus, in the few cases where an equilibrium fails to converge from the first KEP iteration, the initial FBT equilibrium, with standard basis functions, is retained. Taking the last converged solution for each equilibrium, the resulting set of equilibria is then linearly interpolated onto the full RAPTOR time grid and passed in a matrix form to the next RAPTOR simulation. The gradients at the pedestal, $\mu_{ne}$ and $\mu_{Te}$, can also be passed from the previous RAPTOR simulation to the next as a feedforward trace to ease the work of the gradient-based controller. Consistency between transport and equilibrium is obtained when $p'$ and $TT'$ take identical profiles in RAPTOR and FBT, corresponding to a basis function coefficient $\alpha=1$. After iterating over the two codes sequentially, a systematic error on poloidal beta $\Delta\beta_{pol}^{\text{FBT-RAPTOR}}$ and $\Delta\alpha$ of less than $5\%$ remains, due to residual differences in volume and radial resolution. By subtracting $\Delta\beta_{pol}^{\text{FBT-RAPTOR}}$ and coefficient $\alpha$ with their final value, Fig. \ref{fig:convergence-81882} shows that the coupling successfully converges to its final solution within a few iterations, as shown in the example of TCV shot \#81882, where a remaining difference of less than 1\% to the converged values was obtained after only 2 steps. This rapid convergence can be explained by the quasi-unilateral nature of this coupling, as radial transport is only weakly sensitive to changes in geometry, and by the mutual independence of the static equilibrium solutions. It also justifies a posteriori the choice of loose coupling as a valid option. 

The choice of constraining $p'(\psi)$ or $TT'(\psi)$ to match $I_p$ ensures that the shot's initial programming remains unaltered, despite small numerical discrepancies. Contours of the new poloidal flux, represented in Fig. \ref{fig:plasma_flux_comparison_initial_vs_profile}, show a typical correction made to the total flux, $\psi_{total} = \psi$, plasma contribution, $\psi_{plasma}=\mathbb{M}_{xy} \boldsymbol{I_y}$, and coil contribution, $\psi_{coils}=\mathbb{M}_{xa}\boldsymbol{I_a} $, with the new radial profiles. Fluxes are centered around the total flux at the plasma boundary, $\tilde{\psi}=\psi-\psi_{B}$, whose exact value are linked to the radial derivative of the Shafranov shift at the boundary $\Delta'_B\sim \beta_{pol}+l_{i3}/2$ (Eq. 4.21 in \cite{zakharov_equilibrium_1986}). The initial FBT programming is given a very low poloidal beta $\beta_{pol} = 0.07$ and high q on-axis $q_A = 2$, then used as initial condition to the coupling in order to better demonstrate the full procedure. After convergence, FBT-RAPTOR results in a stronger Shafranov shift, with $\beta_{pol}+l_{i3}/2$  being multiplied by $2.2$, and in the PF coils generating a stronger vertical magnetic field in the mid-plane. Keeping the same value of plasma current $I_p$, the new $\beta_{pol}$ coming from RAPTOR and the $q_A$ value corresponding to the new $l_{i3}$ can also be directly imposed as constraints with the usual basis functions, resulting in the same $\beta_{pol}+l_{i3}/2$ as in the coupling. Interestingly, the comparison (Fig. \ref{fig:plasma_flux_comparison_profile_vs_0D}) reveals that the Shafranov shift is not sufficient to explain the change in coil contribution, as the shape of the internal profiles is also responsible for a change in the flux induced by the plasma current, which causes a correction in the order of tens of Amperes in PF coil currents. In both cases, the vacuum vertical field is increased by $10-20\milli\tesla$ around the midplane due to the higher Shafranov shift, thereby leading to a significant decrease and increase of the $E_{4-5}$ and $F_{4-5}$ coil currents respectively.

Fig. \ref{fig:83738:jtor_pprime_ttprime_comparison} shows an example of $j_\phi$, $p'$ and $TT'$ profiles obtained with RAPTOR for different modeling assumptions, with comparison to the simple polynomial profiles initially programmed in FBT. We find that the choice of model for the ion temperature and the quality of the density estimate have a reasonably low impact on the $p'$ and $TT'$ used in Grad-Shafranov for that shot, compared to the profiles used initially.

The new $I_a$ currents calculated by FBT are then included in the usual shot preparation. The ohmic transformer coil current needed to drive the target $I_p$ trajectory and the eddy currents induced by the time variation of ohmic, PF and plasma currents, are computed at the stage from the FBT equilibria. Final values of the feedforward PF coil trajectories are then computed by adding corrections to the pre-computed FBT traces, so as to cancel the stray fields resulting from the transformer and eddy currents. Feedback control during the experiment further adds to the currents modification. The TCV hybrid control system is based on a set of switching matrices responsible for the allocation of coils to the various real-time control tasks, in particular the control of the PF coil currents, vertical and radial displacement, total plasma current and stray field coming from the ohmic (OH) circuit \cite{guizzo_assessment_2024}. Two (external) PF coil combinations are typically selected for the vertical and radial plasma position control, while the internal G coil is used for vertical stability. Overall, the stray field correction of the feedforward traces and real-time feedback control both introduce a systematic gap between the coil currents prepared in the FBT run and the actual currents used during the experiment, which must be taken into account in understanding the results. The latter is discussed in Sec. \ref{Sec:TCV_results}.

\section{Benchmark of the pre-shot prediction of 211 shots}
\label{Sec:PreShotScan}
\begin{figure}[hpb!]
    \centering
    \includegraphics[width=\linewidth]{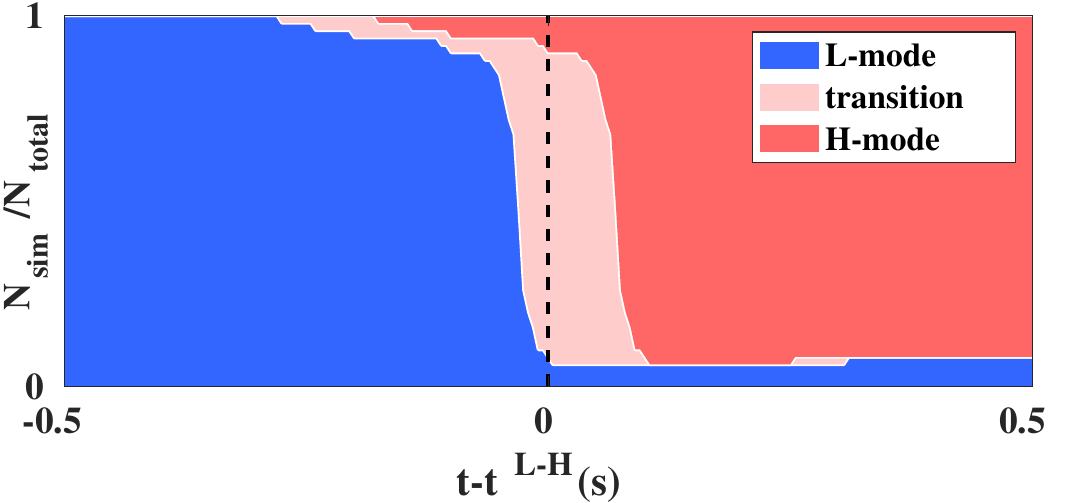}
    \caption{L–H transitions predicted by RAPTOR from the pulse schedule and experimental line-averaged density using Eq. \ref{eq:PsepPLH} for $51$ SN discharges for which a transition was detected from experiments by a confinement state classifier. Each simulation is time-aligned to the detection of either H-mode or D-mode (Dithering) by the classifier at $t^{L-H}$. The vertical axis represents the number of simulations $N_{sim}$ predicted in each confinement state divided by the total number of simulations $N_{tot}$ for a given time slice.}
    \label{fig:PreShotScan:LH_transition}
\end{figure}
\afterpage{
\begin{figure*}[p!]
\centering
\includegraphics[width=0.75\linewidth]{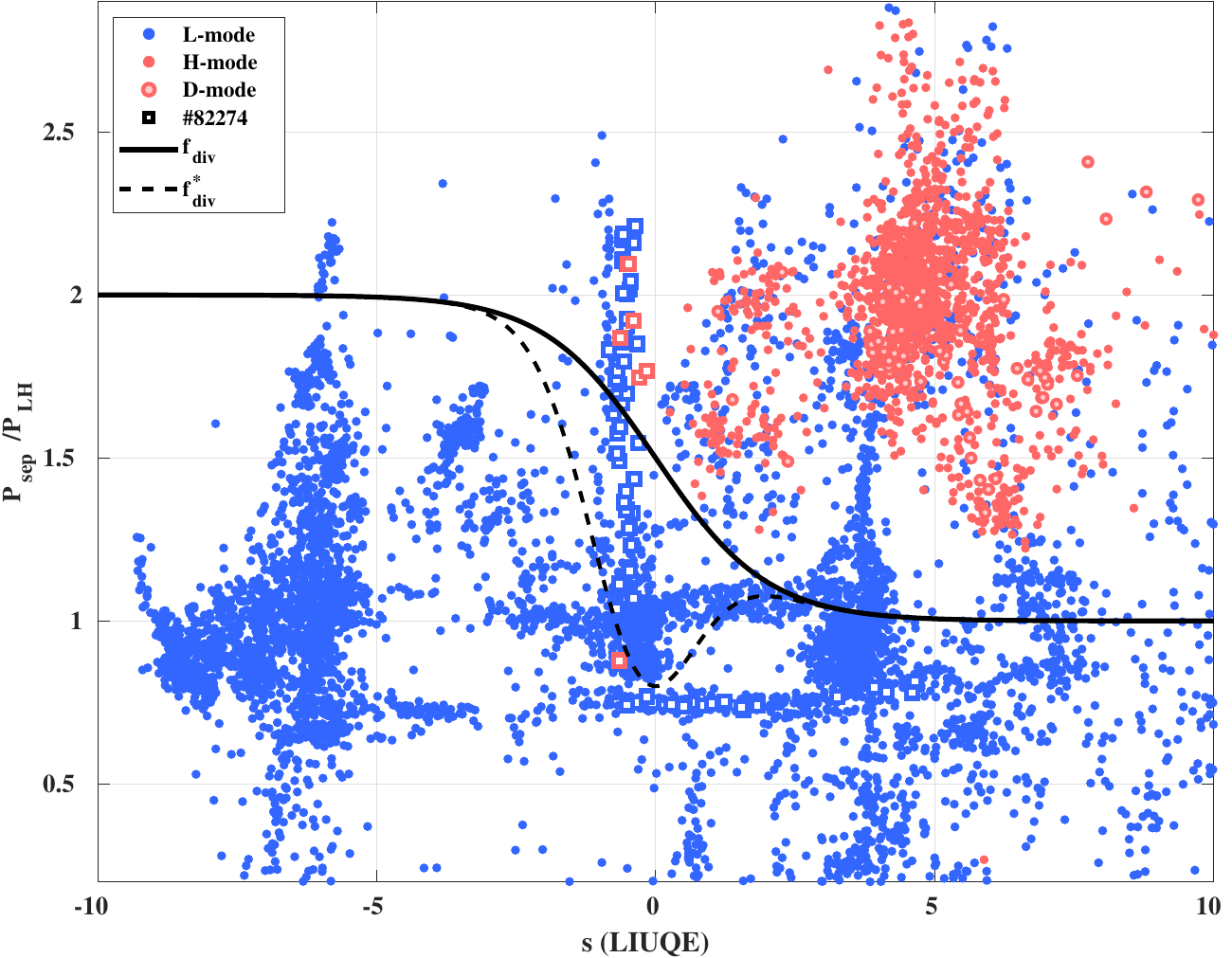}
\vspace{1em}
\caption{Illustration in black of the $f_{div}$ (solid) and $f_{div}^*$ (dashed), functions of $s = \sigma_{fav}|\frac{\delta r_{sep}}{\delta r_{\text{thr}}}|$, Eqs. \ref{eq:alphaLH}-\ref{eq:alphaLH_star}. Each color point corresponds to a time sample from $175$ experiments having a diverted configuration, showing the dependence of the confinement state on the $P_{sep}/P_{LH}$ ratio (computed from experimental data with the standard TCV analysis routines)  and the $s$ value (reconstructed with LIUQE) ; $P_{LH}$, from the Martin scaling with an additional weight at low line-averaged densities ($<3e19m^{-3}$), and $P_{sep}$ estimated assuming a fixed ratio of absorbed NBI power and no core radiation, in accordance with the assumptions made in the RAPTOR simulations. Shot \#82274, a programmed double-null configuration, is highlighted in open squares and further described in Fig. \ref{fig:PreShotScan:DN_NBI_profiles}. The outline of these squares is colored in the same way as the other points, showing the transition to H mode at the end of the NBI ramp, when the $P_{sep}/P_{LH}$ is higher. An alternative version of this plot, accounting for the core radiation in $P_{sep}$, is shown in appendix (Fig. \ref{fig:fiv_curve_rad}). The L, D and H (see legend) phases are marked according to the plasma state classifier \cite{poels_plasma_2025}.}
\label{fig:fiv_curve}
\end{figure*}
}
The methodology presented in Sec.\ref{SubSec:RAPTOR} provides a fast and automatic simulator of the kinetic profiles in the core plasma for full NBI-heated and ohmic discharges. By providing a pre-shot estimate of the internal profiles necessary for solving the Grad-Shafranov equation, Sec.\ref{SubSec:FBT-RAPTOR} showed that these results can be coupled to the FBT inverse equilibrium problem, forming what we refer to as a Kinetic Equilibrium Prediction. In this section, we present the application of this KEP to a wide set of TCV discharges, including $211$ shots in the range $82000-83000$, selected for their duration, availability of Thomson Scattering (TS) data, and excluding discharges with ECRH heating. This shot-base contains a variety of L- and H- mode plasmas in limited and diverted configuration, including NT discharges, and is therefore appropriate for testing the ability of RAPTOR to predict the H-mode phases with the extended Martin scaling law. All the simulations shown in this Section were performed providing the post-shot $n_{e,l}$ from experiments (model E1 in Table \ref{table:EP_models_summary}) as the accuracy of the  $n_{e,l}$ ad-hoc model exposed in Sec. \ref{SubSec:RAPTOR} and later used in the experimental validation in Sec. \ref{Sec:TCV_results} does not yet allow for validation over a large shot database. To compare with our predictions, the actual time of transition from an L- to a Dithering (D-) or H-mode, denoted $t^{LH}$, is determined by a confinement state classifier \cite{poels_plasma_2025} from diagnostic measurements available for each shot, integrated using DEFUSE \cite{pau_modern_2023}. 

\begin{figure}[htb!]
    \centering
    \includegraphics[width=\linewidth]{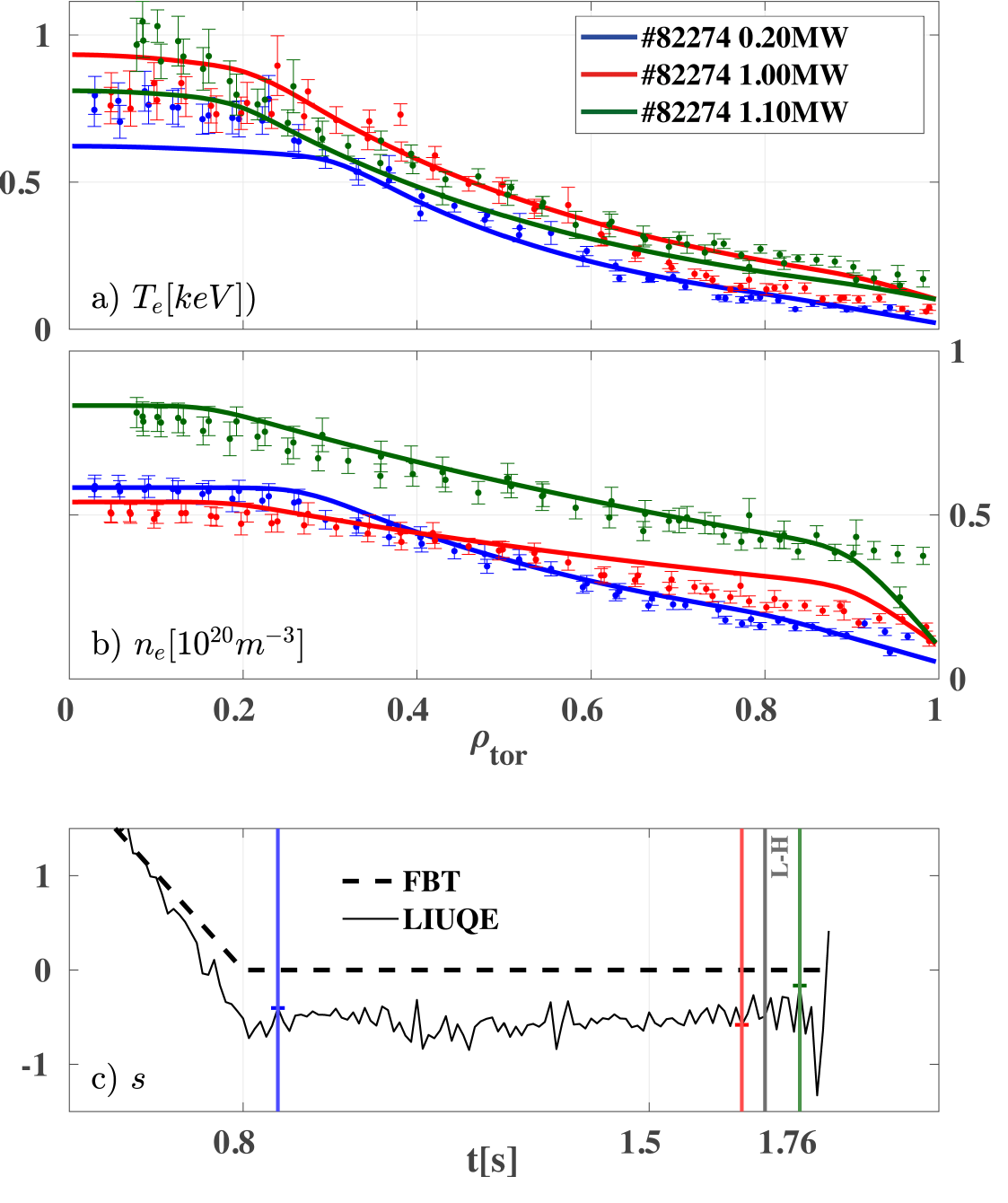}
    \caption{Evolution of $T_e$ and  $n_e$ profiles during the NBI power ramp of (DN) shot \#82274. A delay occurs between the appearance of the H-mode pedestal (a, b) in TS measurements (markers with error-bars) and RAPTOR predicted profiles (model E1, solid line). (c) shows the evolution of $s = \sigma_{fav}|\frac{\delta r_{sep}}{\delta r_{\text{thr}}}|$. This value is programmed to $0$ in FBT (dashed line), to target a DN, LIUQE post-shot (solid line) shows a slightly unfavorable configuration, delaying the access to H-mode. The NBI power ramps linearly from $168\kilo\watt$ at $t=0.8\second$ to $1.12\mega\watt$ at $t=1.8\second$. The experiment only transitions at $t=1.7\second$ (solid vertical grey line). }
    \label{fig:PreShotScan:DN_NBI_profiles}
\end{figure}

\begin{figure*}[htbp!]
\raggedright
\begin{subfigure}[t]{\textwidth}
\centering
\includegraphics[width=0.85\linewidth]{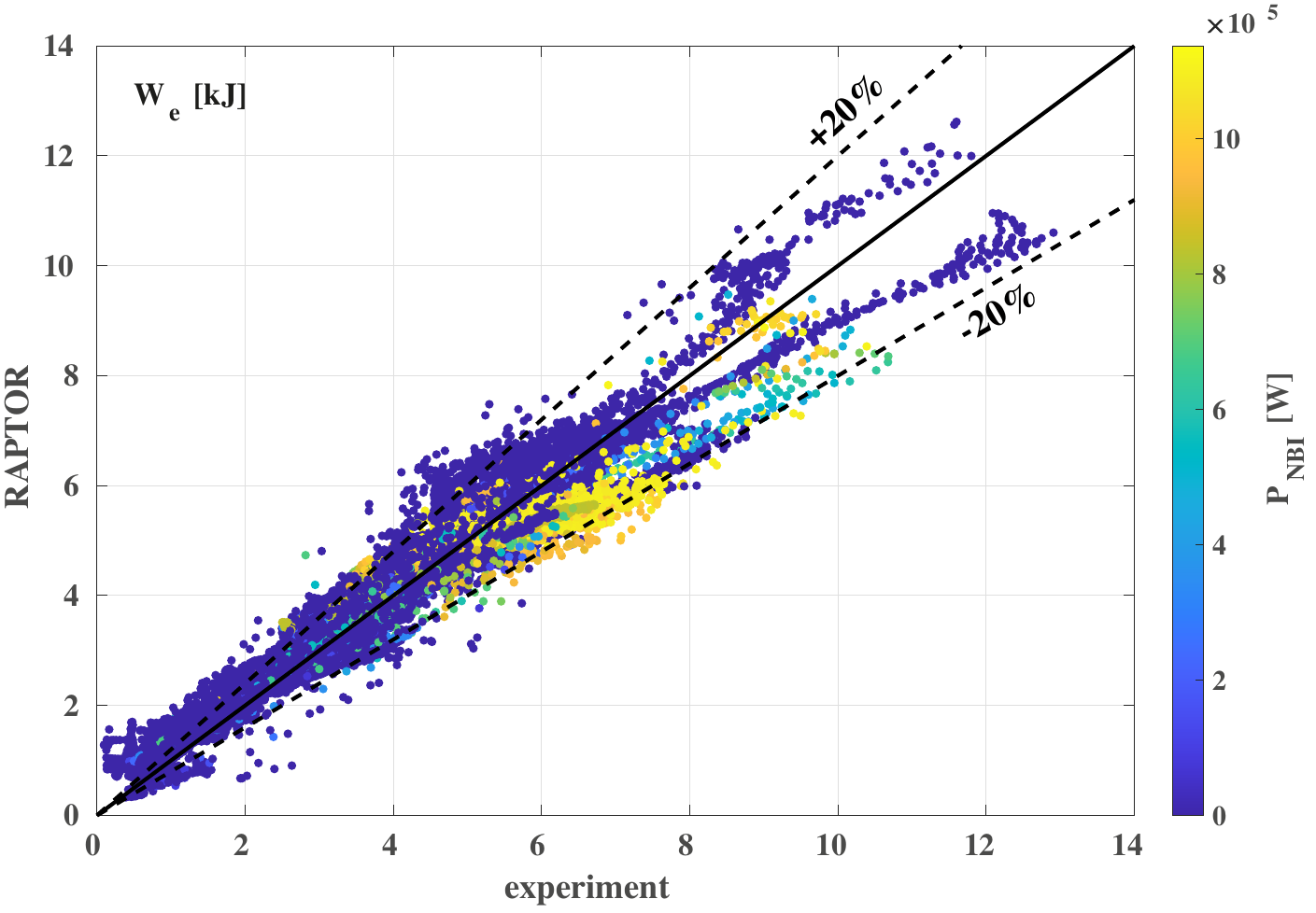}\end{subfigure}
\\[1em]
\begin{subfigure}[t]{\textwidth}
\centering
\includegraphics[width=0.85\linewidth]{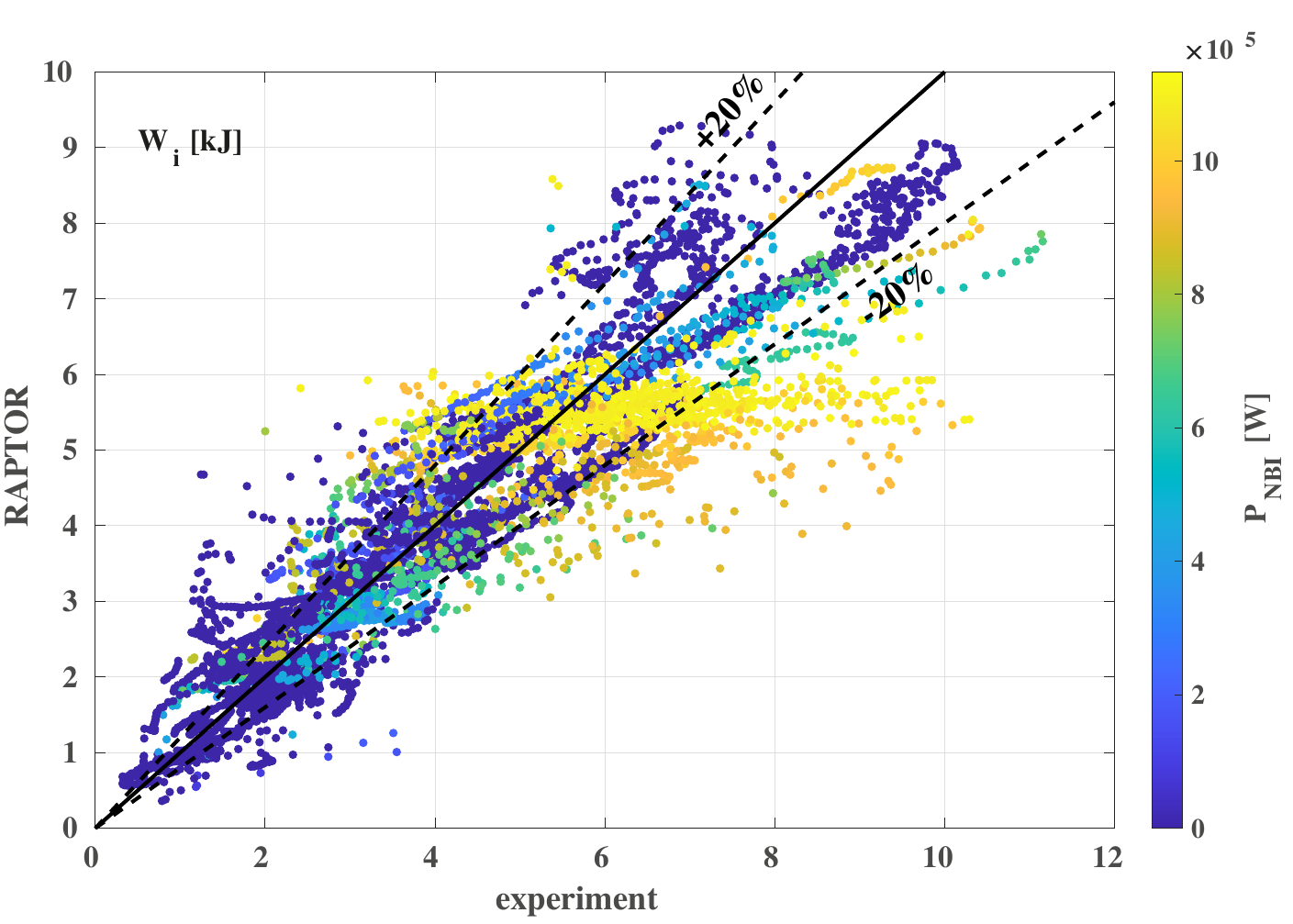}\end{subfigure}
\caption{Comparison of the electron and ion energies predicted by RAPTOR with values reconstructed from experiments for, respectively, $211$ shots with TS and $148$ shots with CXRS measurements, at every measurement time instants. The dashed lines correspond to $W_{e,i}^{\text{experiment}} \pm 20\%$. RAPTOR predictions were made from the pulse schedule, using the line-averaged density from experiments and solving for $T_i$ (model E1 in Table \ref{table:EP_models_summary}). The experimental value of $W_i$ is more uncertain at high power with H-mode and ELM perturbations.}
\label{fig:Wei_prescan_postshot_nel}
\end{figure*}

Fig. \ref{fig:PreShotScan:LH_transition} shows the results of the RAPTOR prediction for $51$ of the $183$ PT shots having an H-mode detected by the classifier at $t = t^{\text{L-H}}$, excluding the DN shot $\#82274$, which is discussed below. For $48$ of these shots, RAPTOR predicts a transition around $t^{\text{L-H}}$ within a $100\milli\second$ time-window. In the remaining $3$ ohmic shots (continuous blue area, Fig. \ref{fig:PreShotScan:LH_transition}), the presence of an H-mode is detected with a low level of confidence and in fact no clear H-mode pedestal is visible in the Thomson Scattering profiles, indicating that the prediction made in RAPTOR, that these shots remain in L-mode, is correct. This good agreement can be explained by a clear restriction of the H-mode access to plasmas with favorable $P_{sep}/P_{LH}$  and $s$ values, as can be seen in Fig. \ref{fig:fiv_curve}. This figure shows experimental values for all of the $183$ PT discharges simulated with the KEP, together with the $f_{div}$ curves presented in Eqs. \ref{eq:alphaLH} (solid, used in the simulations) and \ref{eq:alphaLH_star}. Each non-limited phase of a discharge is represented every $20\milli\second$, with a color corresponding to its confinement state, either detected by the classifier or manually corrected for the $3$ ohmic shots mentioned earlier. While most L-mode plasmas have their $P_{sep}$ below $f_{div}(s)P_{LH}$, some isolated blue points in the upper region can be seen from discharges before or after their transition to H-mode. Denser blue regions can also be seen in the vicinity of the $f_{div}$ curve. These correspond to shots which are marginal and ``resisted'' the H-mode transition for a long period of time despite their favorable $P_{sep}$ and $s$ values. RAPTOR predicted an H-mode in at least one time sample in $25$ of these discharges, including $21$ ohmic shots that stayed in L-mode all along most probably due to the high X-point height of their configuration. This detrimental effect of the X-point height on the H-mode access has been demonstrated experimentally in \cite{andrew_jet_2004, scaggion_eps2012} and could be later added to the extended Martin scaling presented in Sec. \ref{SubSec:RAPTOR}. Among the $4$ remaining NBI-heated discharges for which RAPTOR anticipated an H-mode transition that did not happen or took longer to appear, $3$ come from a study of H-mode access in DN configurations: $\#82272$, $\#82274$ and $\#82275$; with the NBI power ramping from $180\kilo\watt$ to $1.14\mega\watt$ in $1\second$.
These 3 shots are interesting because they were programmed as perfectly CDN in FBT, i.e. with $s=0$, such that the threshold is in between the favorable and unfavorable region, that is in between the dashed and solid black lines in Fig. \ref{fig:fiv_curve}. The 3 discharges behaved very similarly, with a late transition during the NBI ramp and one shot $\#82274$ is highlighted in Fig. \ref{fig:fiv_curve}. First the discharge ended up as an USN with s slightly below 0 ($\simeq-0.5$) as shown in Fig. \ref{fig:PreShotScan:DN_NBI_profiles}c. Since $f_{div}$ is varying rapidly in this region, having a slightly LSN, favorable $s\simeq+0.5$, would have led to a factor 1.25 instead of 1.75. We also show in Fig. \ref{fig:PreShotScan:DN_NBI_profiles} the profiles early in the power ramp and just before and after the transition. RAPTOR predicted a too early transition, as discussed here, and therefore a slightly lower $l_{i3}$. However the predictions with FBT-RAPTOR depend on the accuracy of the shape control in such cases. On the other hand, these shots are the only one to occupy the space between the dashed and solid lines. Since they all transit to H-mode late in the ramp, it shows that TCV does not exhibit a clear favorable effect of DN on the H-mode transition. However more shots are needed to characterize this region more systematically, including cases with varying radiated power and X-point height. 

Finally, an USN NBI-heated discharge in unfavorable configuration and close to the wall, $\#82547$, is located above the $f_{div}$ curve in Fig. \ref{fig:fiv_curve}, where RAPTOR predicts an H-mode, and below, when accounting for the core radiations from bolometer measurements (\ref{fig:fiv_curve_rad}). This shot did not transition to H-mode experimentally, suggesting that further refinement of the model by including impurity transport and line impurity radiation would also help in extending our H-mode prediction capability. 

Comparison of the total energies obtained in experiments and simulations, with and without experimental $n_{e,l}$, and solving for the ion heat equation (models E1 and P1 in Table \ref{table:EP_models_summary}), are presented in Figs. \ref{fig:Wei_prescan_postshot_nel} and \ref{fig:Wei_prescan_preshot_nel}. Results obtained with the electron-to-ion scaling (E2), only solving for the electron temperature and density using the electron confinement factor $H_e^{98(y,2)}$ and $n_{e,l}$, are shown in Fig. \ref{fig:Wei_prescan_postshot_nel_PRETOR}. As shown by the three scans, $W_e$ is predicted within 20\% for essentially all cases, as well as $W_i$ for most cases when $T_i$ is predicted by RAPTOR. The uncertainties are larger for the ion energy content than for the electron energy. The $W_i$ measured by CXRS has larger error bars than TS, particularly in H-mode, as Edge Localised Modes tend to create fluctuations in the passive signal and bias the estimations. Hence RAPTOR is overall clearly predicting within the error bars the full time evolution of the discharges, from ramp-up to ramp-down. The $n_{e,l}$ uncertainty affects the prediction of both electron and ion energies (E1/Fig. \ref{fig:Wei_prescan_postshot_nel} vs. P1/Fig. \ref{fig:Wei_prescan_preshot_nel}), although its impact is more pronounced on specific shots. All three results, solving the four equations taking  $n_{e,l}$ from experiment (E1/Fig. \ref{fig:Wei_prescan_postshot_nel}), solving three equations with $n_{e,l}$ from experiment and electron-to-ion scaling (E2/Fig. \ref{fig:Wei_prescan_postshot_nel_PRETOR}), and the fully predictive model with four equations and predicting $n_{e,l}$ from the pulse schedule (P1/Fig. \ref{fig:Wei_prescan_preshot_nel}), give similar predictions for the L- to H-mode transitions. 

\begin{figure}[htpb!]
    \includegraphics[width=\linewidth]{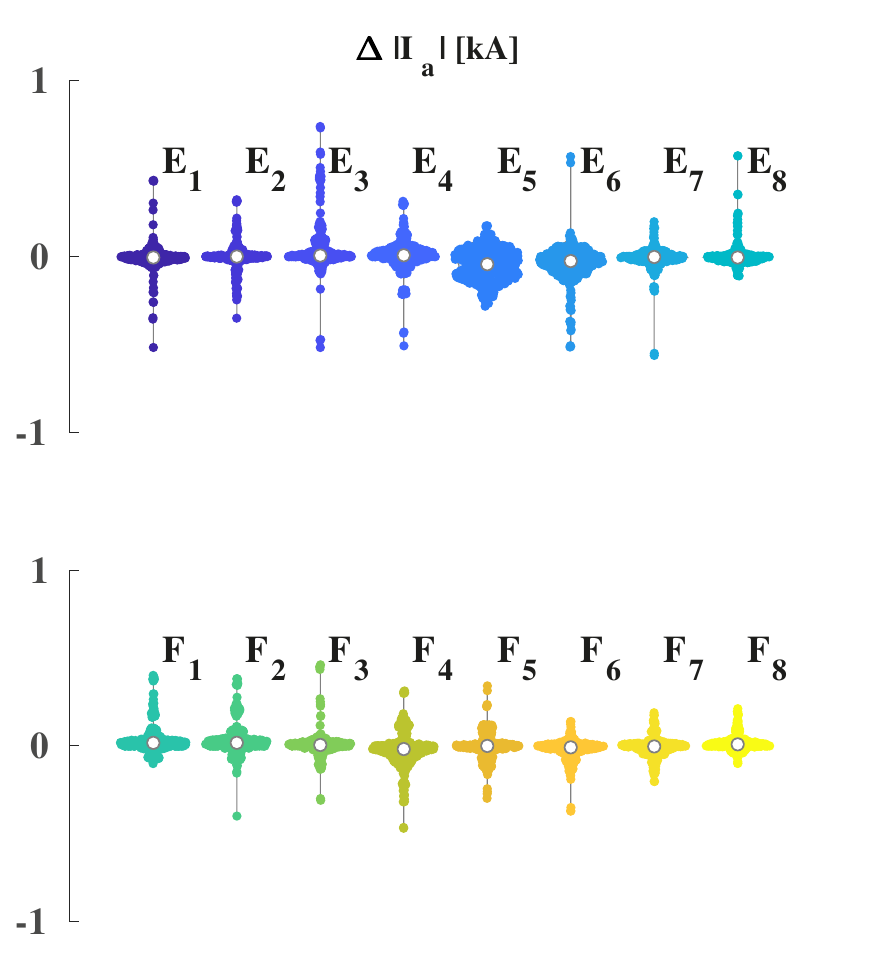}
    \vspace{0.2em}
    \captionsetup{width=\linewidth}
    \caption{Distribution of differences in feedforward PF coil currents (in absolute value) computed by FBT-RAPTOR for various TCV shots, $\Delta |\mathbf{I_a}| = |\mathbf{I_a}^{\textbf{KEP}}|-|\mathbf{I_a^0}|$ for different PF coils, colored as in Fig. \ref{fig:legend_coils_polview}. A cluster of points around $\Delta |\mathbf{I_a}|=0$ indicates that a large number of these corrections remain relatively small (around ten amperes), while larger corrections, on the order of hundreds of amperes, define the upper bound.}
    \label{fig:scan_coils_polview}
\end{figure}

Following the convergence study presented in Sec. \ref{SubSec:FBT-RAPTOR}, three iterations between RAPTOR and FBT were performed for each of the simulations presented above, starting from the initial FBT equilibrium. On average, $6.0\pm 1.5$ minutes of cumulated RAPTOR and FBT simulations were necessary for each iteration of the KEP ($18\pm 4$ minutes per total coupling). We show in Fig. \ref{fig:scan_coils_polview} a distribution of resulting coil currents for times of interest, i.e. for equilibria with plasma currents exceeding $20\%$ of the flat-top value. Since the PF coil currents can take positive or negative values, e.g. depending on the plasma shape or the sign of the plasma current, the difference is calculated from the absolute value, such that positive and negative values of $\Delta |\mathbf{I_a}|$  correspond to an increase or decrease in amplitude respectively. This distribution reveals that the correction made by the KEP occasionally reaches several hundreds of amperes, but is more commonly measured in tens of amperes, depending on the correctness of the initial $\beta_{pol}$ and $q_A$ and the phase in the discharge. Usually, the KEP provides greater corrections in the middle of the shots, where changes in thermal energy can be more important, but the equilibrium remains more steady. It is thus of interest, when preparing for an experiment, to increase the resolution of the FBT-RAPTOR coupling by adding more equilibria in this phase, similarly to what is done in the preparation of experiments presented in Sec. \ref{Sec. TCVResults}. RAPTOR is run only for times with $I_p>50\kilo\ampere$. Simulations converge up to $0.08\pm0.09\second$ in average before the end of their $I_p$ trace, during the ramp-down phase which is often not planned with much care. $100\%$ of FBT equilibria converged up to the last iteration given a converged RAPTOR input profile. The KEP robustness remains therefore mainly limited to the convergence of the latter. All equilibria for which the KEP did not converge from the first iteration (i.e. equilbria for which a converged RAPTOR input profile was missing) are kept with the standard basis functions. 
  
\section{Experimental validation}

\label{Sec. TCVResults}
\label{Sec:TCV_results}
\subsection{Standard H-mode scenario in LSN configuration}
\label{SubSec:81882_shot_results}
\begin{figure*}[htpb!]
  \centering
  \includegraphics[width=\textwidth]{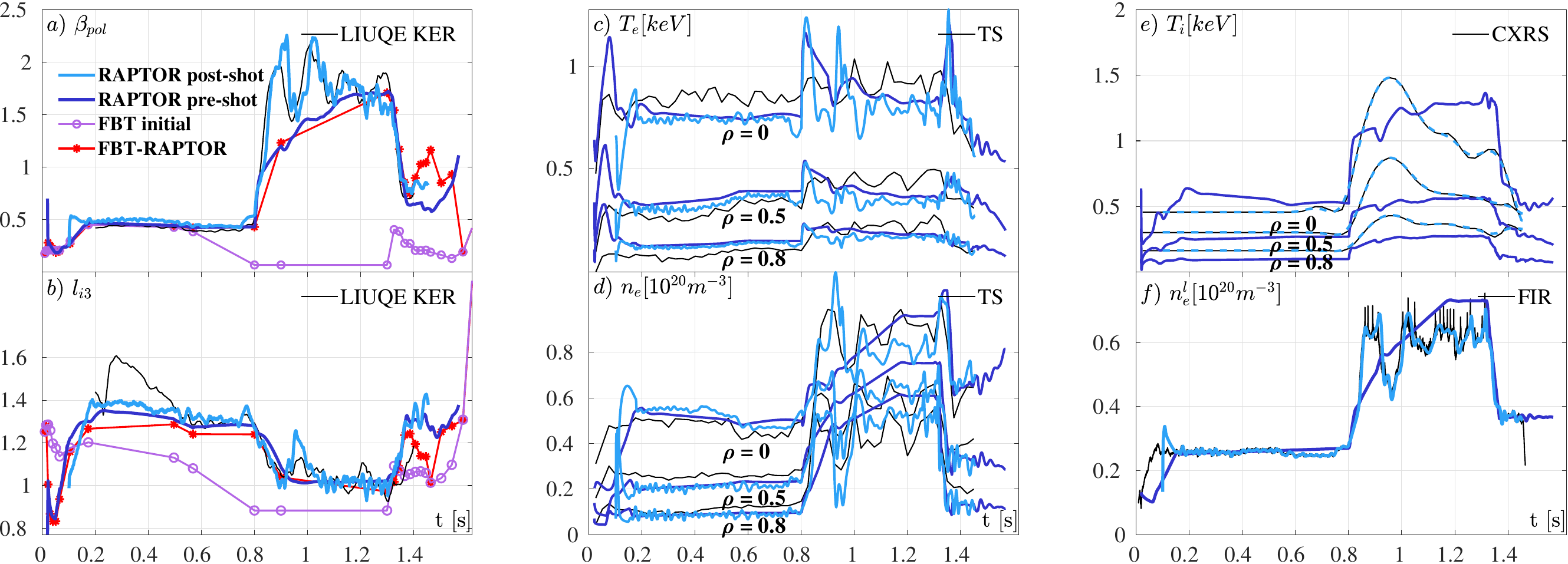}
  \centering
  \vspace{1em}
  \caption{Overview of the KEP performed for TCV shot \#81882. The pre-shot simulation (fully predictive without density measurements, dark blue) is compared to a post-shot simulation (light blue) using the same peaking and confinement factors but taking the line-averaged density and $T_i$ from experiment. TS, CXRS and FIR data of shot \#81882 are displayed in black in (c-f) as well as LIUQE-KER in (a-b). The coupling used an initial FBT run with a low $\beta_{pol}$ and $l_{i3}$ (in purple, (a-b)) used for the preparation of shot \#83738 and the result after 2 iterations (in red) was used to prepare \#83740, both shown in Fig. \ref{fig:83738:shape_comparison}}
  \label{fig:81882:global_view}
\end{figure*}
\begin{figure}[htpb!]
    \centering
    \includegraphics[width=0.6\linewidth]{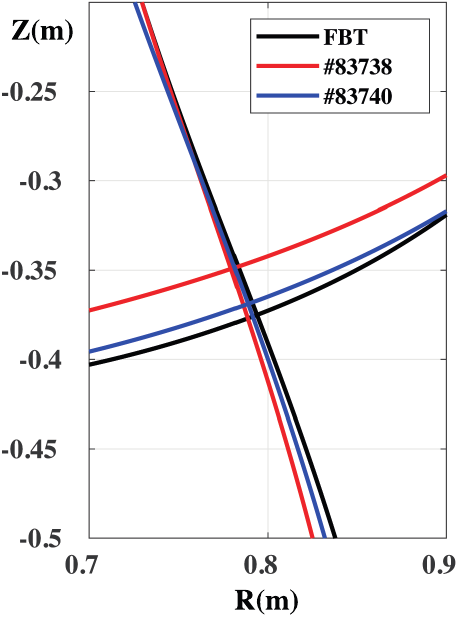}
    \caption{FBT X-point target and LIUQE shape reconstructions obtained in the H-mode phase for shot \#83738, prepared with an initial FBT equilibrium with low-$\beta_{pol}$ and low-$l_{i3}$, and shot \#83740, prepared with the FBT-RAPTOR correction}
    \label{fig:83738:shape_comparison}
\end{figure}

This section presents in more detail the pre-shot simulation and FBT preparation of TCV shot \#81882, using model P2 in Table \ref{table:EP_models_summary}. This shot is a standard scenario with an injection of $1 \mega\watt$ of NBI power in the flat-top phase and a mainly inductive current profile. Starting from a short $100\milli\second$ ramp-up in limited configuration, the plasma becomes diverted at $t=0.4\second$ forming a lower X-point, with ion grad-B drift pointing downwards, in the favorable direction. Shortly after the NBI injection, the power threshold is reached, triggering the transition to H-mode at $t=0.8\second$. The electron confinement quality $H_e^{98}$ is thus slowly increased from $0.4$ to $0.5$, while the target $n_{el}$ is increased up to $n_{el} \simeq 7 \ 10^{19}\meter^{-3}$, as shown in Fig. \ref{fig:81882:global_view}. In this case we used the $T_e/T_i$ scaling law to predict $T_i$. The logarithmic gradient for the electron temperature $\lambda_{Te}$ and the density $\lambda_{ne}$ decrease following Table \ref{table:MS_parameters}, flattening the profiles with the rise of the pedestal and the change of the plasma parameters, modifying the turbulent regime \cite{razumova_main_2008,wagner_understanding_2012}. Comparison with the post-shot simulation (light blue, Fig.  \ref{fig:81882:global_view}), shows that the variations in $n_{e,l}$ and ion temperature are responsible for the small difference in $\beta$ observed in Fig.  \ref{fig:81882:global_view}a.  

The KEP is initialized from a standard FBT programming with polynomial basis functions, low $\beta_{p}$ and high $q_A$ (purple line in Fig. \ref{fig:81882:global_view}) keeping the shape, position and plasma current desired for the shot. Starting from a deliberately degraded estimate of $\beta_{p}$ and $q_{a}$  ($l_{i3}$) allows us to test the robustness of the method and its capability to correct any kind of initial condition. Indeed, consistent values of the internal inductance $l_{i3}$  and poloidal beta $\beta_{pol}$ are obtained between FBT and RAPTOR from this equilibrium after only 2 iterations, changing the flux distribution and the estimated combination of poloidal field coil currents $I_a$ required for maintaining the shape. These new feedforward traces show significant differences during the H-mode phase, reaching several hundred amperes where the change in $\beta_{pol}$ and $l_{i3}$ is more significant. Variations in flux distribution are particularly pronounced in regions of low poloidal field, such as the primary X-point target, even during the least affected phases of the discharge. In these regions, the PF coil flux distribution is influenced as much by differences in $\beta_{pol}$ and $l_{i3}$ as by the shape of $p'$ and $TT'$, though sometimes with a reduced impact on coil currents. We tested the pulse schedule experimentally, with the initial FBT programming (\#83738) and with the new FBT-RAPTOR result (\#83940), as described above. While both shots stayed generally stable with a good plasma shape, a better alignment of the X-point with the FBT target was observed all along the discharge prepared with FBT-RAPTOR, as shown in Fig. \ref{fig:83738:shape_comparison} and confirmed with the MANTIS \cite{perek_mantis_2019} inversion. 

\begin{figure}[hbp!]
  \centering
  \includegraphics[width=0.9\linewidth]{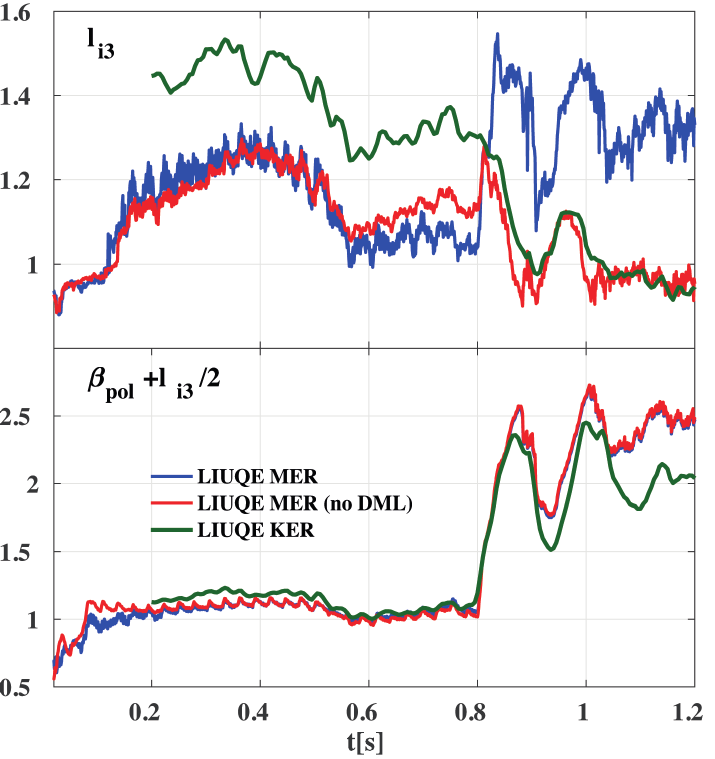}
  \centering
  \caption{Comparison of $l_{i3}$ and $\beta_{pol}$ reconstructed from LIUQE KER and LIUQE MER with and without the DML measurement.}
  \label{fig:83738:betap_li_comparison}
\end{figure}

\paragraph{Comparison between LIUQE MER and KER} is important for validating the KEP predictions, in particular for the evaluation of the internal inductance $l_{i3}$. 
As mentioned notably in \cite{pustovitov_magnetic_2001}, configurations with identical values of $\beta_{pol}+l_{i3}/2$, sharing the same plasma boundary and plasma induced magnetic field on the same boundary, can be attributed to various equilibria with different pressure and current density profiles. In the standard LIUQE Magnetic Equilibrium Reconstruction (MER), which only relies on magnetic measurements to reconstruct the equilibrium of an experiment, the diamagnetic loop (DML) helps distinguishing the $TT'$ from the $p'$ term in the Grad-Shafranov equation \cite{moret_tokamak_2015}, providing an estimate of $l_{i3}$ over $\beta_{pol}$. Fig. \ref{fig:83738:betap_li_comparison}  shows that the ratio using the DML for shot \#81882 differs significantly from the one obtained when performing an interpretative LIUQE-ASTRA KER \cite{carpanese_development_2021} using the kinetic profiles coming from Thomson Scattering and CXRS data, and the current diffusion computed within ASTRA. While the DML constrained MER predicts an increase of internal inductance when passing from L to H-mode (as can be seen by comparing results with and without the DML constraint), the opposite behavior is observed in the interpretative KER, similarly to what is obtained in the pre-shot RAPTOR simulation (Fig. \ref{fig:81882:global_view}), which is coherent with the emergence of an additional bootstrap current density at the H-mode pedestal. Although the true value of the inductance may lie somewhere in between the two results, the idea that the internal inductance decreases with the appearance of this edge bootstrap current term is more realistic. This example illustrates how kinetic-equilibrium coupling, e.g. the RAPTOR-FBT KEP before the shot and a fortiori the ASTRA- or RAPTOR- LIUQE KER after (or during) the shot, can provide additional information beyond results from magnetic measurements alone, by providing a physics-based estimate of $l_{i3}$ and $\beta_{pol}$. 

\begin{figure*}[htpb!]
  \centering
  \includegraphics[width=\textwidth]{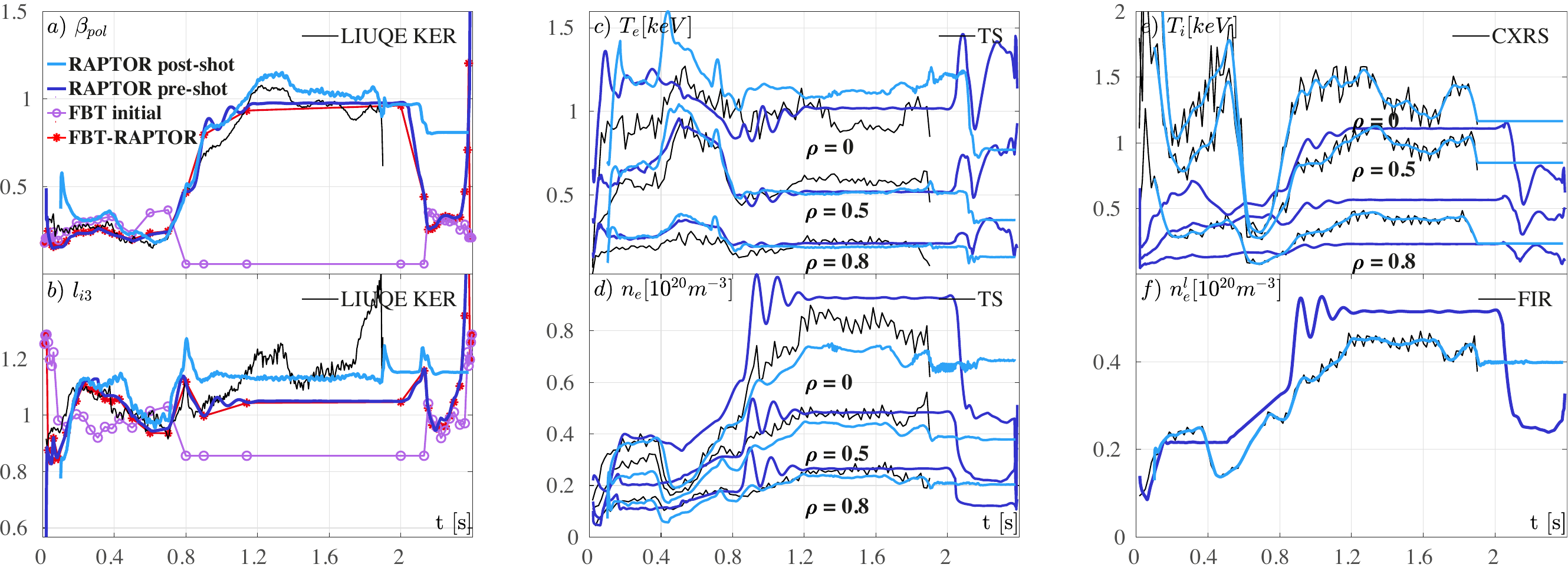}
  \centering
  \vspace{1em}
  \caption{Overview of the KEP performed for TCV shot \#83575. The pre-shot simulation (fully predictive without density measurements, dark blue) is compared to a post-shot simulation (light blue) using the same peaking and confinement factors but taking the line-averaged density and $T_i$ from experiment. TS, CXRS and FIR data of shot \#83575 are displayed in black in (c-f) as well as LIUQE-MER in (a-b). The coupling used an initial FBT run with a low $\beta_{pol}$ and $l_{i3}$ (in purple, a-b) used for the preparation of shot \#83926 and the result after 2 iterations (in red) was used to prepare \#83946, both shown in Fig. \ref{fig:83575_X_points_location}}
  \label{fig:83575:a4_report_paper24}
\end{figure*}
\subsection{Negative-triangularity L-mode scenario with SF divertor target}

\label{SubSec:83575_shot_results}
Fig. \ref{fig:83575:a4_report_paper24} shows the pre-shot simulation (with model P2) of shot \#83575, an upper negative triangularity (NT) L-mode plasma with up to $460\kilo\watt$ of NBI power. The plasma starts in PT-LSN configuration during the current ramp-up and overshoot, then transitions to an upper NT with snowflake divertor (SF) shape at $t = 0.7\second$ \cite{fevrier_iaea_2025} (see Fig. \ref{fig:legend_coils_polview} for the SF shape). 

\begin{figure}[htp!]
\centering
  \includegraphics[width=0.9\linewidth]{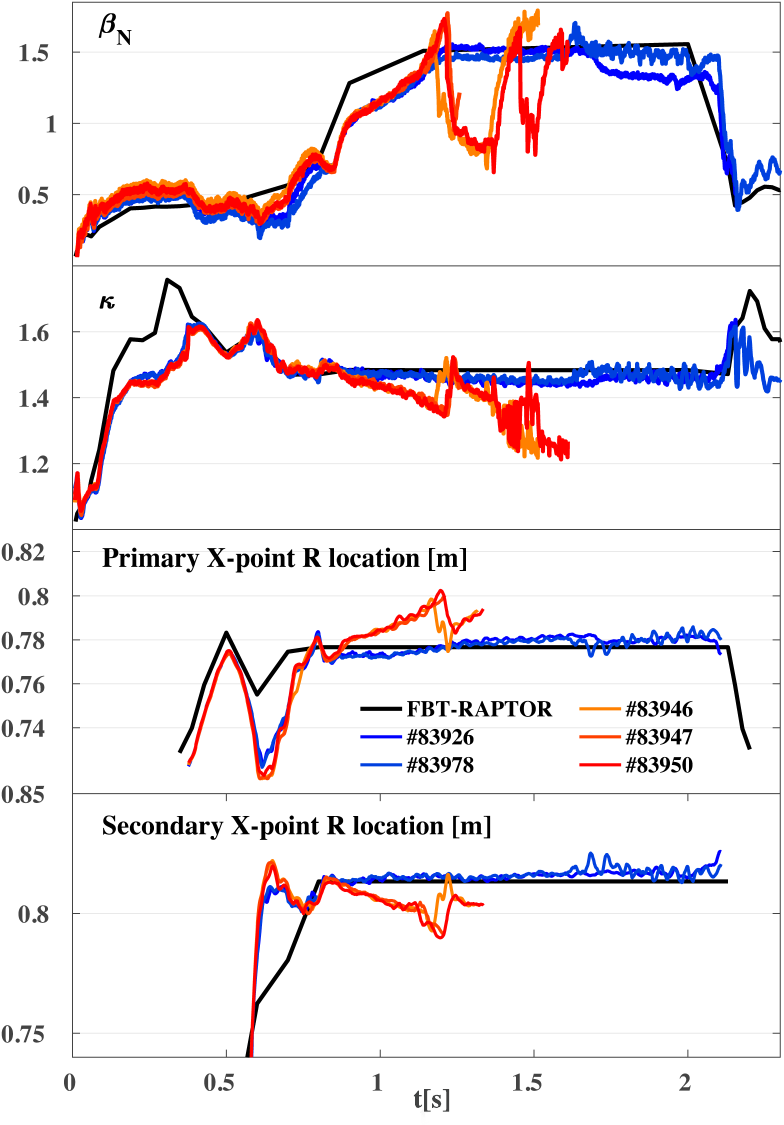}
  \caption{Evolution of $\beta_N$, elongation $\kappa$ and X-point locations for upper NT plasmas with snowflake divertor configuration for a series of shots prepared with the same programming but different equilibrium preparation. Shots \#[83946, 83947, 83950] (in red and orange), prepared with an initial FBT equilibrium with high $\beta_{pol}$ and $l_{i3}$, and shots \#[83926, 83978]  (in blue), prepared with the FBT-RAPTOR corrections.}
\label{fig:83575_X_points_location}
\end{figure}

$H_e^{98(y,2)}=0.6$ is used for predicting the NT phase of the discharge, based on previous discharges, with peaking factors as in Table \ref{table:MS_parameters}. This yields a total $H^{98(y,2)} \sim 1.2-1.3$ using $T_i$ from Eq. \ref{equ:PRETOR}. The line-averaged density is predicted from the pulse schedule, keeping the assumptions made for PT L-mode plasmas. Fig. \ref{fig:83575:a4_report_paper24} shows the resulting pre-shot simulation (dark blue) used for the coupling (in red), compared with the experimental results. The confinement of the discharge out-performed the prediction: the standard TCV post-shot analysis estimates the total confinement factor $H^{98(y,2)}$ between $1.3$ and $1.5$ in the NT-phase, mainly resulting from higher ion temperatures than predicted by the scaling (Fig.\ref{fig:83575:a4_report_paper24}e). A post-shot simulation, repeated (in sky blue) with the same electron confinement parameter ($H_e^{98}=0.6$) and input NBI power, but taking the line-averaged density and ion temperature from experiment, yield a total confinement factor between $1.5$ and $1.6$, resulting from a more conservative absorbed fraction of the NBI power in the simulations. Following the same procedure as in \ref{SubSec:81882_shot_results}, we repeated the scenario experimentally, with the initial FBT programming (purple, Fig. \ref{fig:83575:a4_report_paper24}) for shots \#[83946, 83947, 83950] and with the new FBT-RAPTOR results (red) for shots \#[83926, 83978], providing the new feedforward set of PF coil currents to the shape controller. Negative triangularity plasmas are known to be more difficult to control vertically \cite{pesamosca_improved_2022,marchioni_vertical_2024}, as confirmed by FGE \cite{carpanese_development_2021}, which computes a vertical growth rate of $1389 \hertz$, compared to $115.97\hertz$ for shot $\# 81882$. Fig. \ref{fig:83575_X_points_location} shows that correcting the FBT preparation with the transport prediction enhances the stationarity and controllability of this scenario. The primary and secondary X-points of the snowflake were maintained closer to their respective targets until the end of the discharge, holding the shape remarkably stationary for a longer time than the shots prepared with low-$\beta_{pol}$ and low-$l_{i3}$ but also than the reference shot \#83575. 

\section{Conclusion and outlook}

This paper presents the first results of a Kinetic Equilibrium Prediction for TCV full discharges, using RAPTOR for fast transport prediction and FBT for the inverse equilibrium, investigating the role of kinetic profiles in the feedforward coils programming and our ability to predict kinetic and current density profiles before the shot. These discharges cover ramp-up, flat top and ramp-down phases, limited and diverted shape evolutions, L- and H-mode phases, ohmic and NBI heating. TCV has 16 PF coils controlled by 16 independent power supplies, which allow to generate a large number of plasma shapes. The Martin scaling law \cite{martin_power_2008} has been included in a new RAPTOR module and adapted to limited and unfavorable diverted configurations, automatically predicting the confinement mode and transitions during the simulation (Sec. \ref{SubSec:RAPTOR}). This H-mode prediction shows promising results in most of the cases studied (Sec. \ref{Sec:PreShotScan}), and is mainly limited by the limited access to H-mode in high X-point height configurations and the role of radiation losses. The specific cases of near double-null shapes also need to be studied further and can provide insights to the L-H threshold evolution from favorable to unfavorable configurations. Despite the generally low level of core radiation in TCV plasma, modeling of impurity transport and line radiation using the already implemented ADAS cooling factors database \cite{ADAS,maget_healing_2022,van_mulders_scenario_2023} is envisioned in the future. It will improve the estimation of the core profiles and power crossing the separatrix $P_{sep}$. The gradient-based model \cite{teplukhina_simulation_2017} has been extended to Negative Triangularity L-mode transport (Sec. \ref{SubSec:RAPTOR}). NT fixed parameters, which reproduce best the set of simulated discharges (listed in Table \ref{table:MS_parameters}), corroborate the H-mode like performances observed in NT plasmas \cite{mariani_negative_2024}. Similarly, simulation of the ion heat transport has been included in the gradient-based model for the first time with RAPTOR over a wide range of discharges, and compared to a simple electron-to-ion data-driven scaling law (Sec. \ref{SubSec:RAPTOR}). We conclude that similar results can be obtained within the CXRS error-bars, provided that a total confinement factor is given that corresponds to the expected performance of the shot. In Section \ref{Sec:PreShotScan}, we show that, for a wide range of discharges, simulating from the pulse schedule and from the experimental $n_{e,l}$ with a default fixed set of parameters (Table \ref{table:MS_parameters}), we predict the total ion and electron energy content within a $20\%$ error-bars, with a good prediction of the L to H mode transition and of the overall kinetic profiles. The estimation of the line-averaged density from the gas programming has proven to be a crucial element of predict-first modeling. Its improvement, e.g. by using more advanced regression strategies or edge-to-core modeling, appears to be an important step toward the automatization and generalization of the prediction strategies presented in this work. However, we present a simple model for TCV that proved sufficient to design and test experimentally our predictions (Sec.\ref{Sec:TCV_results}), and therefore allow validating the pulse schedule.

In Section \ref{SubSec:FBT-RAPTOR}, we show that internal profiles ($p'$, $TT'$) obtained from the RAPTOR pre-shot simulations can help inform the Grad-Shafranov equation in the FBT inverse equilibrium solver. FBT is used before a discharge for preparing the plasma shape and feedforward PF coil currents and is usually parametrized with simple assumptions on the core profiles. We show that the FBT-RAPTOR coupling enables the integration of a more complex plasma dynamics while converging to a final solution within a few iterations and minutes, with a negligible systematic error between the two codes, hence a self-consistent solution is obtained with this loose coupling. 
We also demonstrate that not only an accurate time evolution of $\beta_p$ and $l_i$ is important, as expected, but the profiles of $p'$ and $TT'$ (of $j_\phi(R,Z)$) also play a significant role for accurate control.
Section \ref{Sec:TCV_results} demonstrates the implementation of this new result in the TCV shot preparation system, by giving the new feedforward currents to the hybrid shape controller, facilitating the feedback control during the discharge. Experiments show that getting a better prediction of $\beta_{pol}$ and $l_{i3}$ and of the profiles can help improving the alignment of the shape and avoid vertical instabilities during the shot. In particular, the new coupling helped to better control a snowflake configuration with an upper NT plasma, maintaining the primary and secondary X-points close to their target until the end of the discharge. This kind of shape, known to be more difficult to control, proved to be more sensitive to the FBT-RAPTOR coupling than a standard PT-LSN configuration, for which the principle effect was a better alignment of the X-point target. Such physics-based results can then be used to inform a more advanced shape control model by providing better feedfoward coil currents and dynamical profiles of the plasma.

We chose to have a loose coupling between FBT and RAPTOR, i.e. to compute equilibrium and transport separately, in order to facilitate its portability to the existing TCV shot preparation system, and given that transport calculations are only weakly dependent on the equilibrium. However, a tight coupling  which would run RAPTOR, FBT and fast models for auxiliary heating and current-drive systems, e.g. TORBEAM \cite{poli_torbeam_2018} for Electron Cyclotron Resonance Heating, RABBIT \cite{weiland_rabbit_2018} for Neutral Beam Injection, at different time steps of a single run, could be of interest, as changes in heating and current sources have a stronger influence on the predicted plasma profiles. The gradient-based model, which assumes three regions of flat, stiff and pedestal gradients over the whole radius, can later be amended to include local changes in confinement, e.g. to include effects of the magnetic shear usually observed in advanced scenarios (as already done in other RAPTOR transport models \cite{felici_real-time_2011}), or turbulence surrogate models like QLKNN \cite{Fransson_QLKNN_new_2023} or TGLF-NN \cite{cao_tglf-winn_2026}. Finally, with the OH circuit equations already implemented in FBT, their integration into the KEP represents a natural extension of this work. This would allow the joint optimization of the plasma current induction together with the control of the plasma shape, thereby reducing the gap between the predicted and effective PF coil currents and enhancing both the precision and impact of the KEP in the preparation of TCV discharges.

\section*{Data availability}
The simulation results presented in this article will be made available in a publicly accessible Zenodo repository upon acceptance of the paper. They may be accessed and reused in accordance with the conditions specified therein.

\section*{Acknowledgments}
This work has been carried out within the framework of the EUROfusion Consortium, partially funded by the European Union via the Euratom Research and Training Programme (Grant Agreement No 101052200 — EUROfusion). The Swiss contribution to this work has been funded in part by the Swiss State Secretariat for Education, Research and Innovation (SERI). Views and opinions expressed are however those of the author(s) only and do not necessarily reflect those of the European Union, the European Commission or SERI. Neither the European Union nor the European Commission nor SERI can be held responsible for them. This work was supported in part by the Swiss National Science Foundation.

\section*{ORCID iDs}
C. E. Contré \orcidlink{0009-0008-3246-1860} \href{https://orcid.org/0009-0008-3246-1860}{0009-0008-3246-1860} \\
A. Merle \orcidlink{0000-0003-1831-5644}\href{https://orcid.org/0000-0003-1831-5644}{0000-0003-1831-5644} \\
O. Sauter \orcidlink{0000-0002-0099-6675} \href{https://orcid.org/0000-0002-0099-6675}{0000-0002-0099-6675} \\
S. Van Mulders \orcidlink{0000-0003-3184-3361} \href{https://orcid.org/0000-0003-3184-3361}{0000-0003-3184-3361} \\
 R. Coosemans \orcidlink{0000-0001-8110-3156}\href{https://orcid.org/0000-0001-8110-3156}{0000-0001-8110-3156} \\
 G. Durr-Legoupil-Nicoud
 \orcidlink{0009-0002-5956-6482}\href{https://orcid.org/0009-0002-5956-6482}{0009-0002-5956-6482} \\
F. Felici \orcidlink{0000-0001-7585-376X} \href{https://orcid.org/0000-0001-7585-376X}{0000-0001-7585-376X} \\
O. Février \orcidlink{0000-0002-9290-7413}\href{https://orcid.org/0000-0002-9290-7413}{0000-0002-9290-7413} \\
C. Heiss \orcidlink{https://orcid.org/0009-0005-9712-0643}\href{https://orcid.org/0009-0005-9712-0643}{0009-0005-9712-0643}\\
B. Labit \orcidlink{0000-0002-0751-8182}\href{https://orcid.org/0000-0002-0751-8182}{0000-0002-0751-8182} \\
A. Pau \orcidlink{0000-0002-7122-3346}\href{https://orcid.org/0000-0002-7122-3346}{0000-0002-7122-3346} \\
Y. Poels \orcidlink{0000-0002-4071-4855}\href{https://orcid.org/0000-0002-4071-4855}{0000-0002-4071-4855} \\
C. Venturini \orcidlink{0009-0005-9873-1171}\href{https://orcid.org/0009-0005-9873-1171}{0009-0005-9873-1171} \\
B. Vincent \orcidlink{0000-0001-5420-6002}\href{https://orcid.org/0000-0001-5420-6002}{0000-0001-5420-6002}

\section*{Author contributions}
\textbf{C. E. Contré}: Writing – review \& editing, Writing – original draft, Investigation, Software, Conceptualization, Methodology, Visualization, Formal analysis;
\textbf{A. Merle}: Writing – review \& editing, Investigation, Methodology, Software;
\textbf{O. Sauter}: Writing – review \& editing, Supervision, Investigation, Conceptualization, Methodology, Formal analysis, Validation, Software;
\textbf{S. Van Mulders}: Writing – review \& editing, Conceptualization, Formal analysis, Software;
\textbf{R. Coosemans}: Writing – review \& editing, Formal analysis (ASTRA KER);
\textbf{G. Durr-Legoupil-Nicoud}: Resources (NT-SF scenario);
\textbf{F. Felici}: Writing – review \& editing, Supervision, Software;
\textbf{O. Février}: Resources (NT-SF scenario);
\textbf{C. Heiss}: Visualization;
\textbf{B. Labit }: Resources (PT-LSN scenario), Data curation (QCE);
\textbf{A. Pau}: Writing – review \& editing, Formal analysis (confinement state classification);
\textbf{Y. Poels}: Formal analysis (confinement state classification);
\textbf{C. Venturini}: Formal analysis (confinement state classification);
\textbf{B. Vincent}: Writing – review \& editing, Data curation (CXRS);

\bibliographystyle{iopart-num-noURL-max3authors-keepdots}
\bibliography{references}

\clearpage
\onecolumn
\appendix
\counterwithin{figure}{section}
\renewcommand{\thefigure}{\thesection.\arabic{figure}}
\setcounter{section}{0}
\section{Additional figures}
\begin{figure*}[htpb!]
\centering
\includegraphics[width=0.65\linewidth]{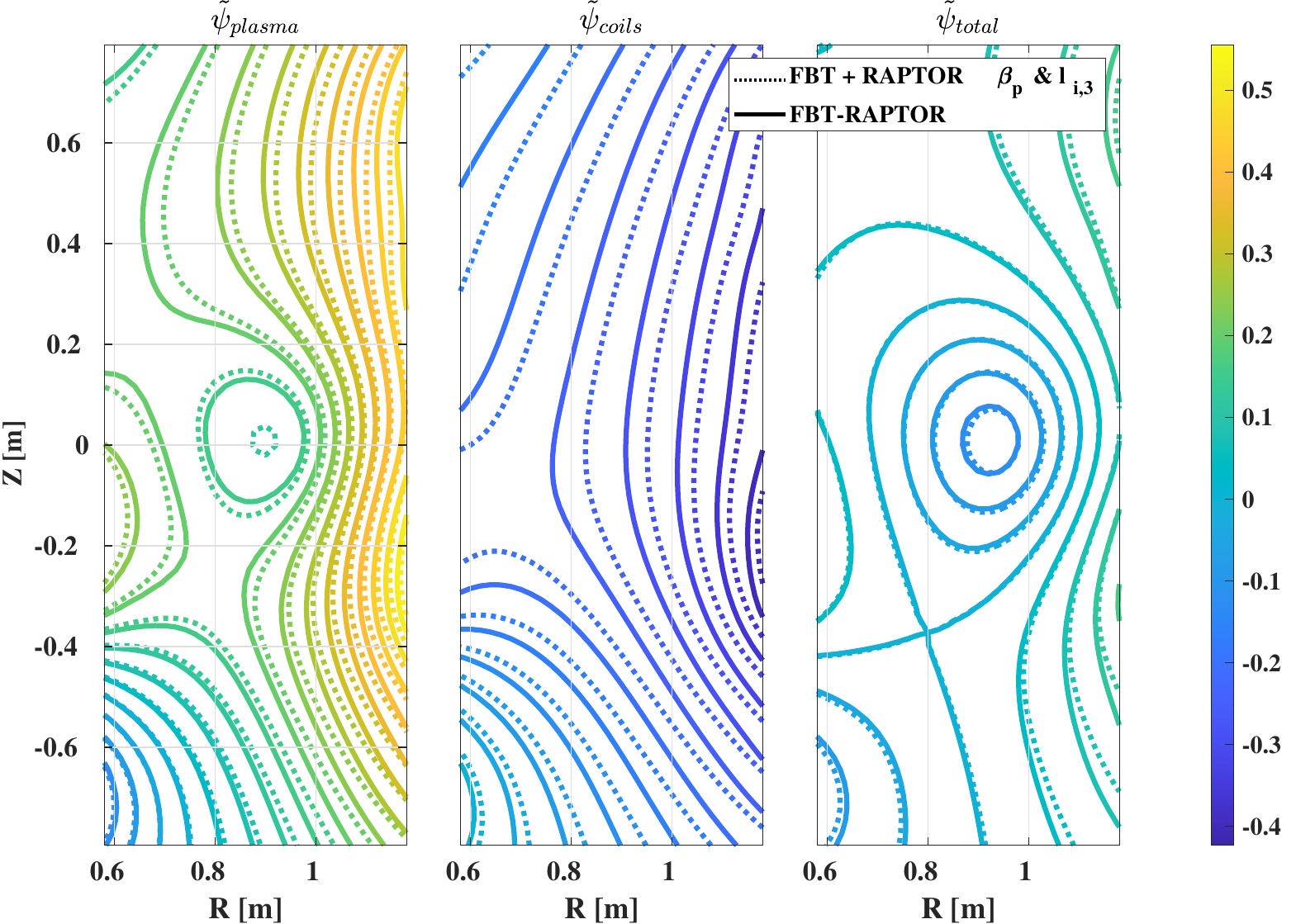}
\caption{Plasma and coils contribution to the total poloidal flux in the H-mode phase ($t=1.3\second$) of \#81882 centered around the total flux at the plasma boundary, $\tilde{\psi}=\psi-\psi_{B}$. The full profile coupling (solid line) is compared to constraining $\beta_{pol}$ and $q_A$ from RAPTOR in the initial FBT calculation and keeping the same value of $I_p$ (in dotted line), showing non negligible differences in plasma flux surfaces and PF coils flux generation}
\label{fig:plasma_flux_comparison_profile_vs_0D}
\end{figure*}

\begin{figure*}[htpb!]
\centering
\includegraphics[width=0.6\linewidth]{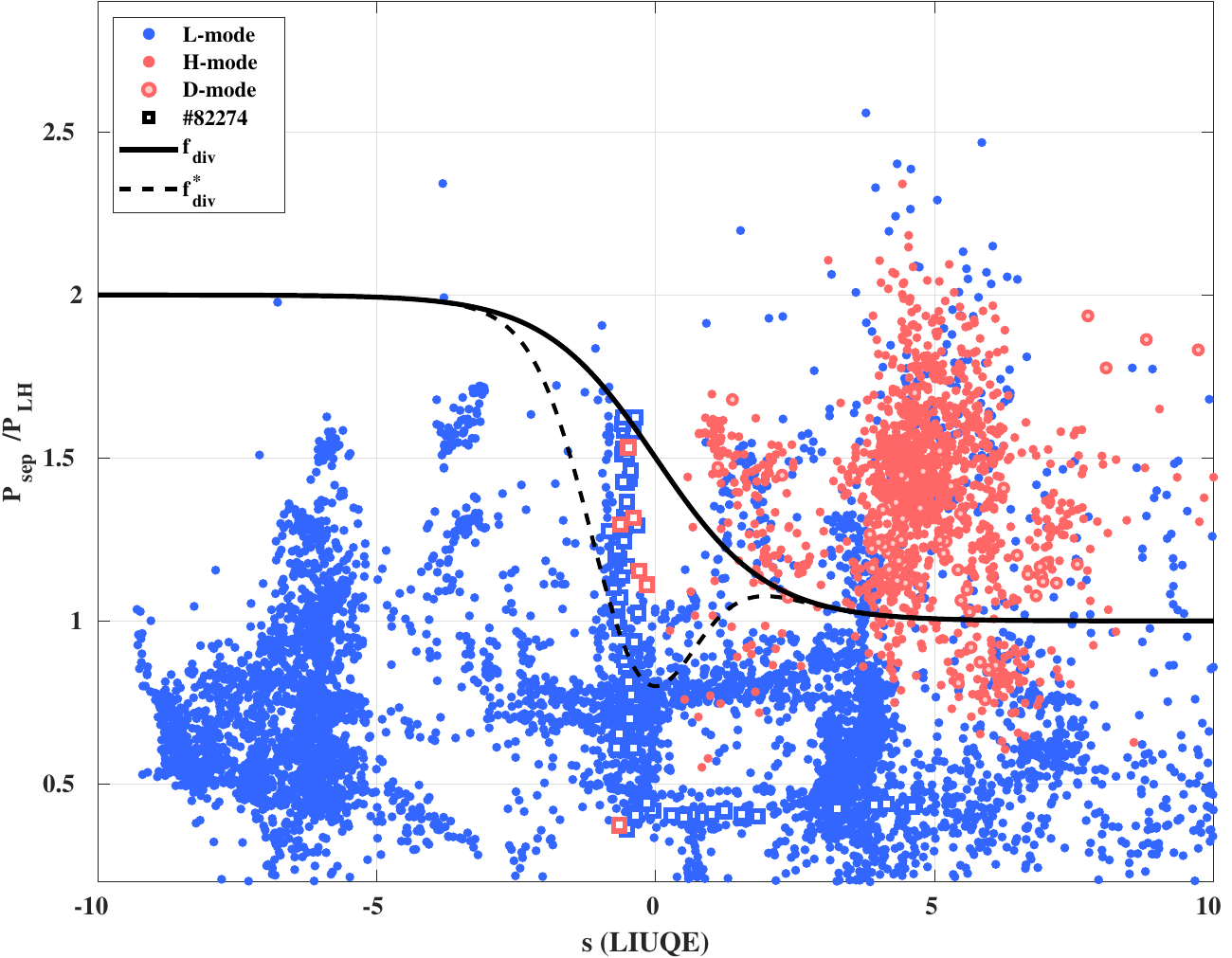}
\vspace{1em}
\caption{Alternative version of Figure \ref{fig:fiv_curve} accounting for core radiation in the computation of $P_{sep} = P_{oh} + P_{aux} - P_{rad, core}$, from the RADCAM tomographic inversion \cite{sheikh_radcamradiation_2022}. Each point corresponds to a time sample from $175$ experiments having a diverted configuration, showing the dependence of the confinement state on the $P_{sep}/P_{LH}$ ratio and the $s$ value (reconstructed with LIUQE). The $f_{div}$ function is the same as used in simulations and in Fig. \ref{fig:fiv_curve}.}
\label{fig:fiv_curve_rad}
\end{figure*}

\begin{figure*}[p]
\raggedright
\begin{subfigure}[t]{\textwidth}
\centering
\includegraphics[width=0.9\linewidth]{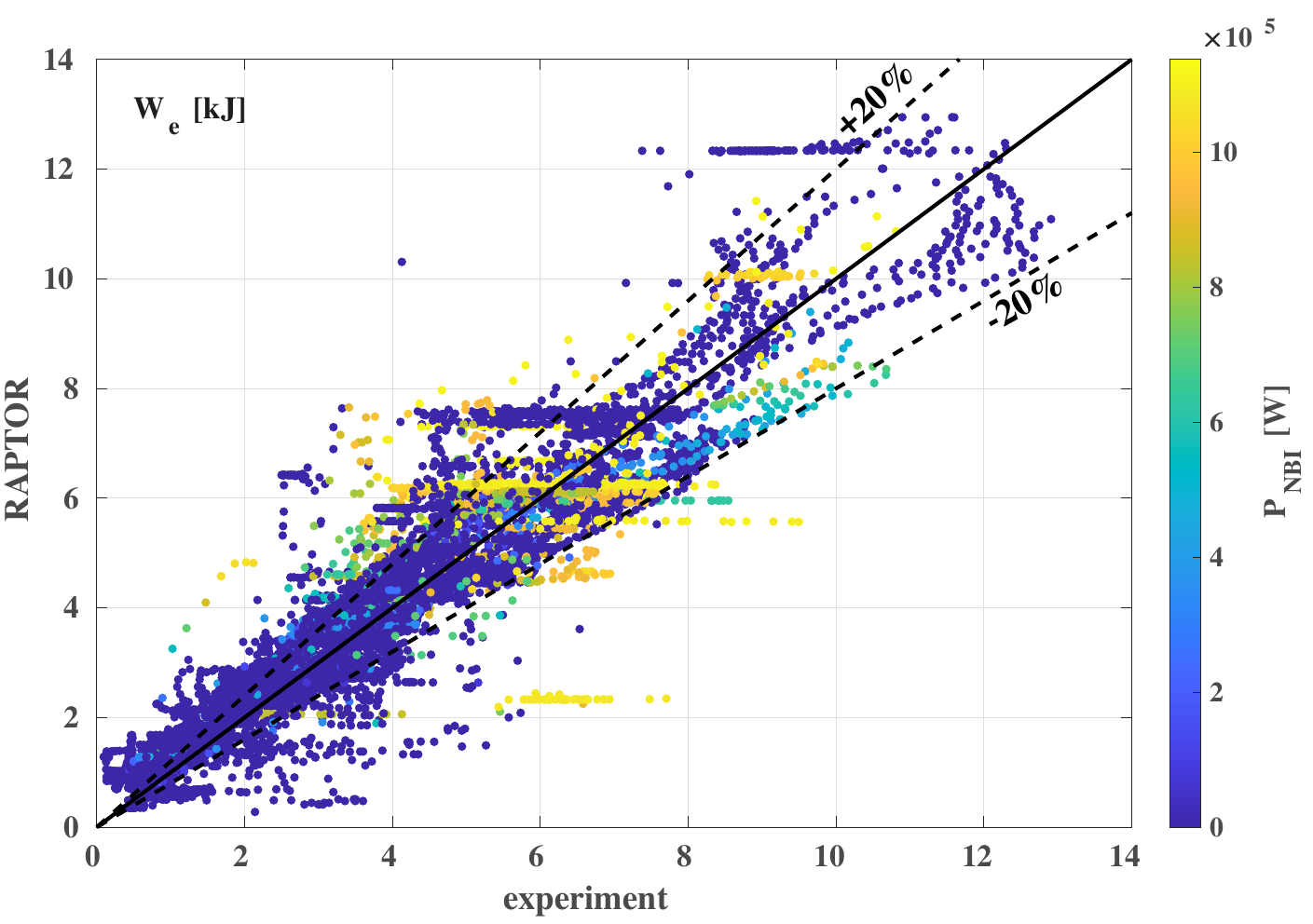}\end{subfigure}
\\[1em]
\begin{subfigure}[t]{\textwidth}
\centering
\includegraphics[width=0.9\linewidth]{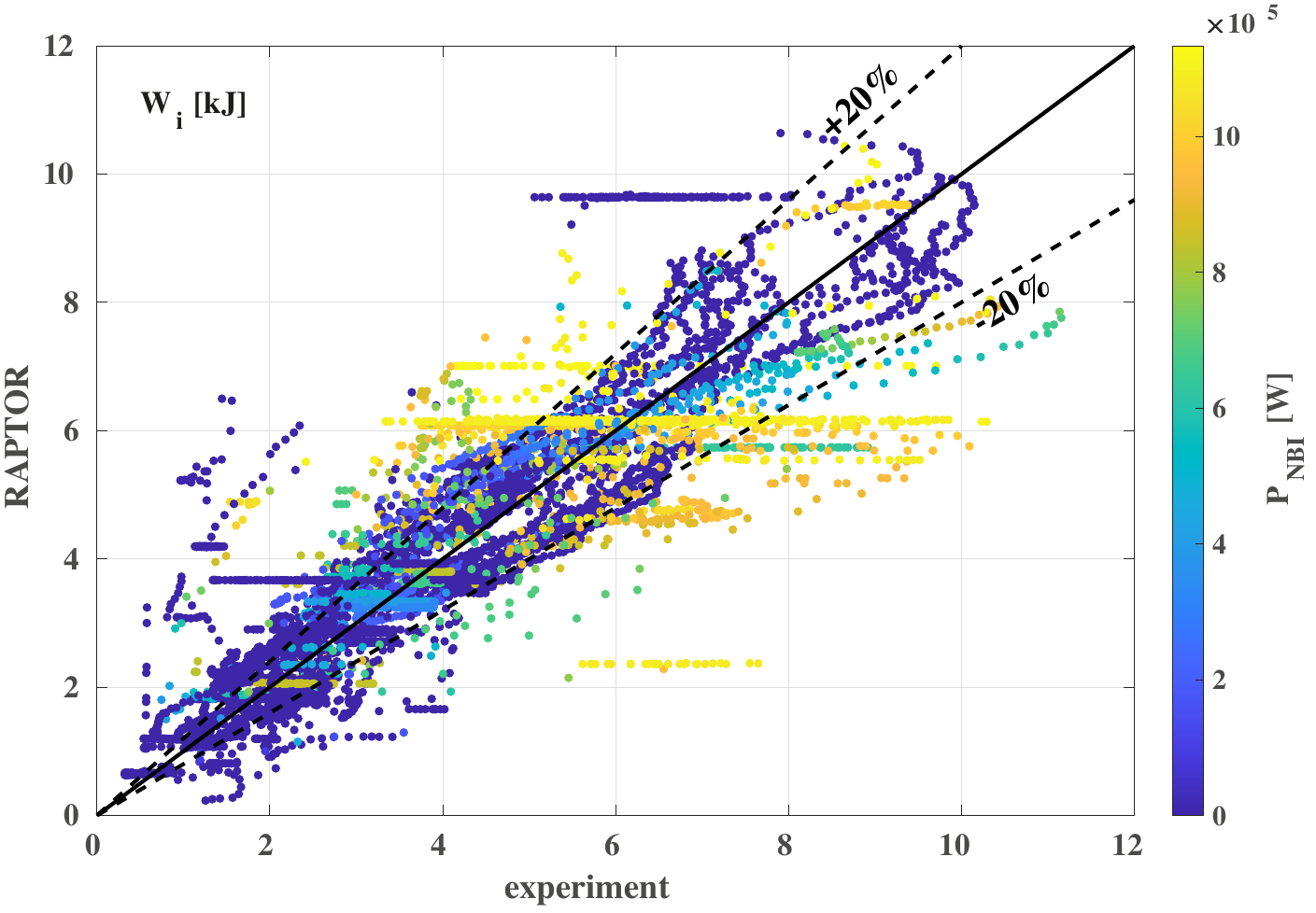}\end{subfigure}

\caption{Comparison of $W_e$ and $W_i$ predicted by RAPTOR with values reconstructed from experiments for, respectively, $211$ shots with TS and $160$ shots with CXRS measurements, at every measurement time instants. The dashed lines correspond to $W_{e,i}^{\text{experiment}} \pm 20\%$. RAPTOR predictions were made from only the pulse schedule, predicting $n_{e,l}$ from the gas references and solving for $T_i$ (model P1 in Table \ref{table:EP_models_summary}).}
\label{fig:Wei_prescan_preshot_nel}
\end{figure*}

\begin{figure*}[p]
\raggedright
  \begin{subfigure}[t]{\textwidth}
  \centering
  \includegraphics[width=0.9\linewidth]{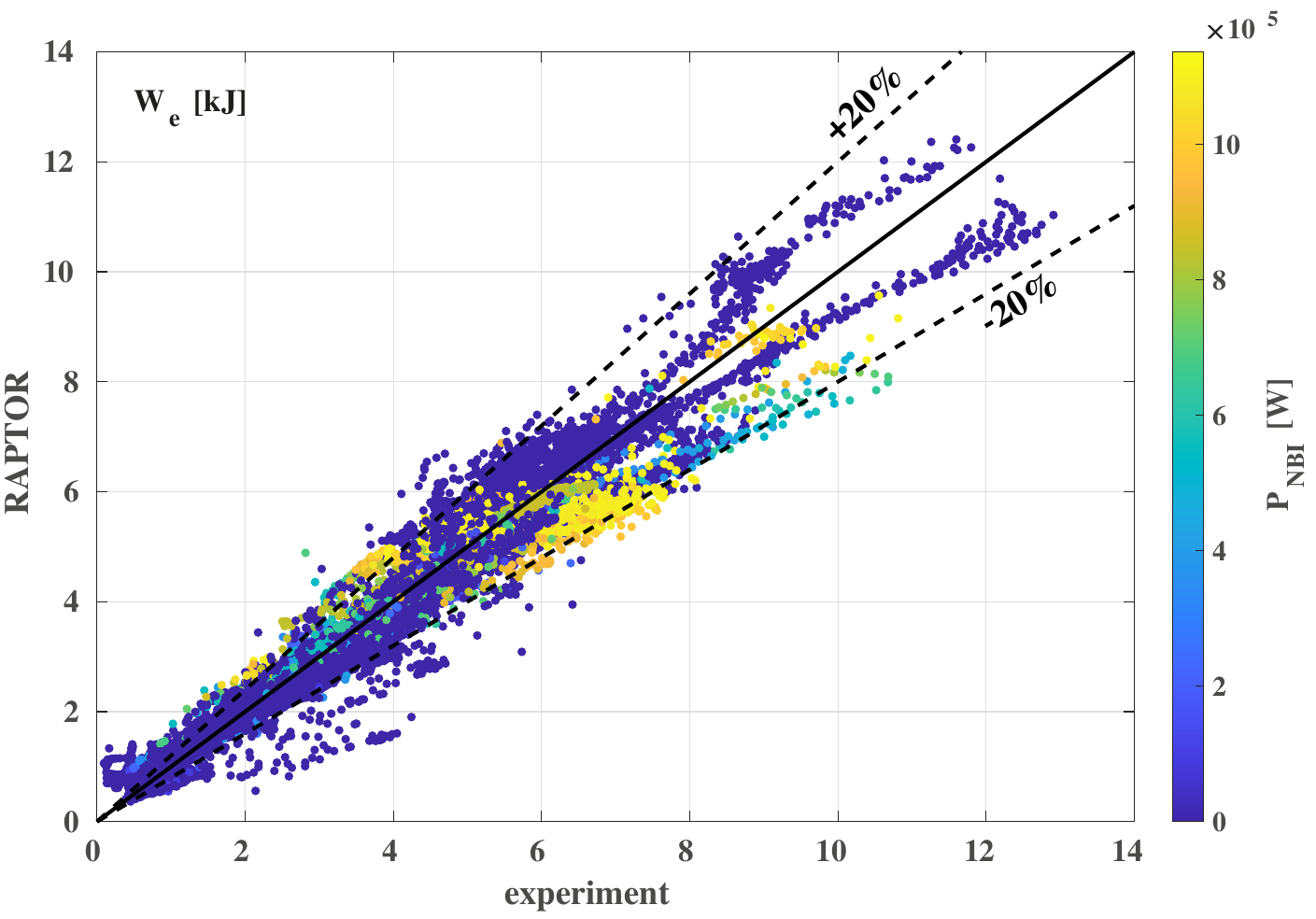}\end{subfigure}
\\[1em]
\begin{subfigure}[t]{\textwidth}
\centering
\includegraphics[width=0.9\linewidth]{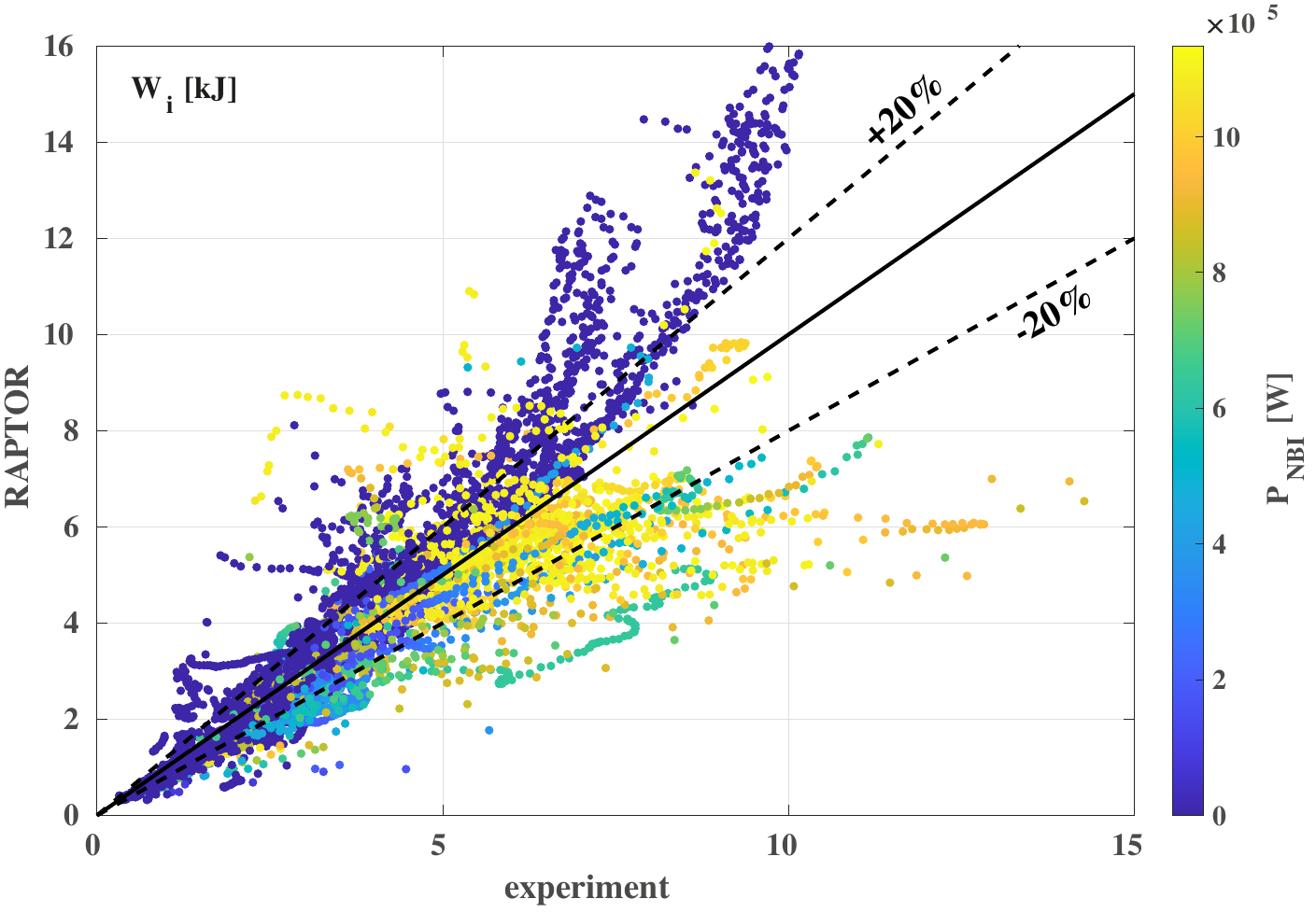}
\end{subfigure}
\caption{Comparison of $W_e$ and $W_i$ predicted by RAPTOR with values reconstructed from experiments for, respectively, $206$ shots with TS and $156$ shots with CXRS measurements, at every measurement time instants. The dashed lines correspond to $W_{e,i}^{\text{experiment}} \pm 20\%$. RAPTOR predictions were made from the pulse schedule, using the experimental $n_{e,l}$ and the electron-to-ion scaling in Eq. \ref{equ:PRETOR} (model E2 in Table \ref{table:EP_models_summary}).}
\label{fig:Wei_prescan_postshot_nel_PRETOR}
\end{figure*}

This appendix provides additional figures to complement the ones presented in the discussion. Figure \ref{fig:plasma_flux_comparison_profile_vs_0D} shows the impact of providing the full RAPTOR profiles to FBT to the poloidal magnetic flux generated by the main plasma and external PF coils, compared to simply constraining the FBT profiles with RAPTOR predicted 0D quantities ($\beta_{pol}$ and $q_A$). Figure \ref{fig:fiv_curve_rad} shows the effect of including the core radiation on the $P_{sep}/P_{LH}$ values for each of the PT diverted shots presented in the benchmark (as compared to Fig. \ref{fig:fiv_curve} without $ P_{rad, core}$). Finally, Figures \ref{fig:Wei_prescan_preshot_nel} and \ref{fig:Wei_prescan_postshot_nel_PRETOR} show the predicted total thermal electron and ion energies energies against their experimental values for models P1 and E2, as compared to Fig. \ref{fig:Wei_prescan_postshot_nel} (model E1 in Table \ref{table:EP_models_summary}).

\end{document}